\documentclass[reprint,notitlepage,aps,pre,10pt,amsmath,amssymb,showkeys,floatfix]{revtex4-2}
\usepackage[utf8x]{inputenc}
\usepackage{graphicx}
\usepackage{epstopdf}
\usepackage{epsfig}
\usepackage{xcolor}
\usepackage{cancel}
\usepackage{ulem}
\usepackage{xcolor}
 \usepackage{url}

\DeclareMathOperator{\sech}{sech}
\newcommand{\dbar}{{\mathchar '26\mkern -11.5mu\mathrm{d}}}

\begin{document}

\title{A Microcanonical Inflection Point Analysis via Parametric Curves and its Relation to the  Zeros of the Partition Function} 

\author{J. C. S. Rocha}
  \email{jcsrocha@ufop.edu.br}
  \affiliation{Departamento de Física, Universidade Federal de Ouro Preto - UFOP, Ouro Preto, Minas Gerais, Brasil.}
  
\author{R. A. Dias}
  \email{rodrigo.dias@ufjf.br}
  \affiliation{Departamento de Física, Universidade Federal de Juiz de Fora - UFJF, Minas Gerais, Brasil.}
  
\author{B. V. Costa}
   \thanks{Retired Professor}
  \email{bvc@fisica.ufmg.br}
  \affiliation{Departamento de Física, Universidade Federal de Minas Gerais - UFMG, Minas Gerais, Brasil.}
\date{\today}

\begin{abstract}
In statistical physics, phase transitions are arguably among the most extensively studied phenomena. In the computational approach to this field, the development of algorithms capable of estimating entropy across the entire energy spectrum in a single execution has highlighted the efficacy of microcanonical inflection point analysis, while Fisher’s zeros technique has re-emerged as a powerful methodology for investigating these phenomena. 
 This paper presents an alternative protocol for analyzing phase transitions using a
 parametrization of the entropy function in the microcanonical ensemble. We also provide a clear demonstration of the relation of the linear pattern of the Fisher's zeros on the complex inverse temperature map (a circle in the complex $x=e^{-\beta \varepsilon}$ map) with the order of the transition, showing that the latent heat is inversely related to the distance between the zeros.  We study various model systems, including the Lennard-Jones cluster, the Ising, the XY, and the Zeeman models. By examining the behavior of thermodynamic quantities such as entropy and its derivatives in the microcanonical ensemble, we identify key features—such as loops and discontinuities in parametric curves—which signal phase transitions' presence and nature. This approach can facilitate the classification of phase transitions across various physical systems.
\end{abstract}

% keywords can be removed
\keywords{Statistical Physics, Thermodynamics, Phase transition, Entropy, Fisher zeroes}

\maketitle

%%%%%%%%%%%%%%%%%%%%%%%%%%%%%%%%%%%%%%%%%%%%%%%%%%%%%%%%%%%%%%%%%%%%%%%%%%%%%%%%%%%%%%%%%%%%%%%%%%%%%%%%%%
\section{\label{intro} Introduction}
%%%%%%%%%%%%%%%%%%%%%%%%%%%%%%%%%%%%%%%%%%%%%%%%%%%%%%%%%%%%%%%%%%%%%%%%%%%%%%%%%%%%%%%%%%%%%%%%%%%%%%%%%%

Phase transitions are ubiquitous in nature, manifesting in phenomena such as the boiling of water and the demagnetization of a magnet. These transitions are among the best-understood emergent phenomena, where the collective behavior of the components results in substantial changes in the macroscopic properties of a system~\cite{kivelson2024statistical}. Understanding phase transitions is essential for fields like materials science, soft and condensed matter physics, and even cosmology~\cite{Fultz_2020, riste2012phase,linde}. 

While the boiling of water involves the coexistence of clearly distinct liquid and vapor phases, this feature is absent in the phase transition of a magnet's spontaneous magnetization. The latter does not exhibit a discernible contrast between ferromagnetic and paramagnetic states at the transition point. Furthermore, it is typically associated with power-law divergences of thermodynamic quantities, such as magnetic susceptibility and specific heat, a phenomenon known as critical behavior. According to P. Ehrenfest, these phase transitions are classified as first-order and second-order, respectively~\cite{Sauer}. This classification is based on the lowest derivative of the free energy that is discontinuous or infinite at the transition point. However, theoretical and experimental findings challenged the completeness of this scheme. Lars Onsager's solution of the two-dimensional Ising model, for example, demonstrated that the free energy's derivative diverges logarithmically near the critical point~\cite{Onsager}, a behavior not fully captured by Ehrenfest's original framework. Consequently, by the 1970s, a more generalized binary classification had achieved prominence. Analogous to Ehrenfest’s scheme, it categorized phase transitions as either first-order, denoting those involving latent heat, or second-order (also called continuous) otherwise~\cite{jaeger}.  

The 1970s also witnessed the description of the so-called Berezinskii-Kosterlitz-Thouless (BKT) transition in two-dimensional systems. Classified as an infinite-order transition in the Ehrenfest scheme, this transition is driven by the behavior of topological defects and has been observed in specific magnetic systems, in superconducting, and superfluid films~\cite{KBT_nobel}. In this work, we employ the framework of statistical physics to contribute to the development of an analytical scheme for studying these phenomena.

From the statistical physics perspective, the thermodynamics of an isolated system can be described by the microcanonical ensemble.  The fundamental equation in this ensemble is the entropy, given by
\begin{equation}
S(E) = k_B\ln{\Omega(E)},
\label{PlanckBoltzmann}
\end{equation}
where $\Omega(E)$ represents the number of states with energy $E$, and  $k_B$ denotes the Boltzmann constant~\cite{Planck}. This expression encapsulates all the essential information required to describe the system.
According to the axiomatic approach of Callen~\cite{Callen}, entropy is a strictly monotonically increasing and concave function in regions where the system attains positive temperatures, and strictly decreasing and concave where negative temperatures are possible.  Thermodynamic systems with an unbounded phase space are incapable of reaching negative temperatures~\cite{ramsey}. The presence of a convex region in the entropy function indicates thermodynamic instability. Specifically, a change in the curvature of $S(E)$ signals a first-order phase transition~\cite{gross2001microcanonical,gross}.
 
Canonically, large fluctuations in certain physical quantities, such as the average energy, typically occur near phase transitions, as indicated by singularities in the specific heat. In contrast, since the temperature remains constant throughout the phase transition, it is expected that the inverse microcanonical temperature,
 \begin{equation}
\bar{\beta}(E) = \frac{1}{\bar{T}} = \left( \frac{\partial S}{ \partial E}\right)_{\{X\}},
\label{beta_e}
\end{equation}
 minimally responds to changes in energy. Therefore, a first-order phase transition is signaled by the existence of a minimum in $\bar{\beta}(E)$. The overbar denotes that we are specifically referring to microcanonical temperatures. It is important to note that, in the thermodynamic limit,  $\bar{T}$ converges to the regular temperature, $T$, usually associated with a heat bath.  Moreover, $\{X\}= V, N, M,\cdots $ represents a set of independent extensive parameters such as volume, $V$, number of particles, $N$, magnetization, $M$, and so forth, that characterize the thermodynamic system. It is worth emphasizing that energy varies more smoothly as a function of $\bar{\beta}$ than of $\beta$.

Furthermore, the occurrence of a convex intruder in the entropy during a first-order transition results in multiple energy values sharing the same $\bar{\beta}$.  Since entropy can be expressed as a function of temperature~\cite{zemansky}, we explore an alternative representation of entropy within the microcanonical ensemble by utilizing the inverse temperature, $\bar{\beta}$, as a parameter. Specifically, we propose to solve Wq.~(\ref{beta_e}) for $E$, yielding $E = E(\bar{\beta})$. This expression is then intended for substitution into Eq.~(\ref{PlanckBoltzmann}) to obtain  $S(\bar{\beta})$.  However, the transformation from $E$ to $\bar{\beta}$ 
is not bijective, as multiple energy values may correspond to identical $\bar{\beta}$ values.  Consequently, the resulting function $S(\bar{\beta})$ fails to satisfy the condition of domain uniqueness of a function in the unstable region. This feature can be used as a criterion for identifying phase transitions\footnote{See Chapter 9 in ref.~\cite{Callen}, more specifically Sections 9-4 and 9-5}.

In the case of a system in thermal contact with a heat bath, its statistical description is usually given by the canonical ensemble. The partition function, $Z$, is the fundamental quantity in this context.  Mathematically, this function can be interpreted as the Laplace transform of \( \Omega(E) \), i.e.,
\begin{equation}
Z(\mathcal{B}) = \int \Omega(E)e^{-\mathcal{B} E} \mathrm{d}E,
\label{Z_total}
\end{equation}
where \(\mathcal{B} =  \beta + i  \tau\) represents a complex inverse temperature, with \(\beta = 1/k_BT \) denoting the regular canonical inverse temperature~\cite{luscombe2021statistical,rodeo}. The canonical ensemble is connected to thermodynamics through the Helmholtz free energy, given by
\begin{equation}
F(\mathcal{B})= -\frac{1}{\mathcal{B}} \ln{Z(\mathcal{B})}.
\label{Helmholtz}
\end{equation}
Although the complex temperature lacks physical meaning, the analytic continuation of the free energy can reveal phase transitions at the limit of $\tau\to 0$~\cite{Lee_Yang}. Specifically, in the Fisher's zeros analysis~\cite{Fisher}, phase transitions are identified by the points where  the zeros pinch in on the real axis, mathematically:
\begin{equation}
\lim_{\begin{smallmatrix}
      N\to \infty \\
      \tau \to 0 
\end{smallmatrix}} Z(\mathcal{B}) \to 0.
\label{Zcondition}
\end{equation}
Examination of Eqs. (\ref{Helmholtz}) and (\ref{Zcondition}) reveals that the zeros of the partition function correspond to the nonanalytic points of the free energy, which manifest as discontinuities and singularities characteristic of phase transitions.

 The main goal of this manuscript is to parametrize the entropy function in the context of the microcanonical ensemble and analyze its region of non concavity behavior. 
Additionally, we aim to demonstrate the relationship between Fisher's zeros maps and these curves. The proposed study is applied to well-known models with first and second-order transitions, as well as to models presenting BKT transitions and no transitions as a matter of comparison. The motivation for studying various types of transitions extends beyond illustrating the proposed analysis; it also has the potential to be used in the development of classifiers within an Artificial Intelligence framework designed to categorize phase transitions\cite{Carrasquilla2017}.

This paper is structured as follows. In Sec.~\ref{fishersZeros}, we present the fundamental concepts of the Fisher zeros analysis. Following that, in Sec.~\ref{fishersZeros1order}, we provide an alternative and simplified demonstration of the connection between the pattern of zeros maps and the unstable region of the entropy, as previously reported~\cite{Rocha_2024}. In the demonstration provided here, we show that the latent heat can be determined by the distance between the zeros in the pattern associated with the first-order transition. In Sec.~\ref{microcanonical}, we discuss the microcanonical inflection point analysis and introduce its parametric formulation.
Section~\ref{results} presents results for various models: the Lennard-Jones cluster in Sec.~\ref{resultsLJ} as a prototype of a first-order transition, the Ising model in Sec.~\ref{resultsIsing} as an example of a second-order transition, the XY model in Sec.~\ref{resultsXY} to study the BKT transition, and the Zeeman model in Sec.~\ref{resultsZeeman} as a case with no transition. Section~\ref{conclusions} outlines our conclusions and offers perspectives for future work.

%%%%%%%%%%%%%%%%%%%%%%%%%%%%%%%%%%%%%%%%%%%%%%%%%%%%%%%%%%%%%%%%%%%%%%%%%%%%%%%%%%%%%%%%%%%%%%%%%%%%%%%%%%
\section{\label{method} Methodology}
%%%%%%%%%%%%%%%%%%%%%%%%%%%%%%%%%%%%%%%%%%%%%%%%%%%%%%%%%%%%%%%%%%%%%%%%%%%%%%%%%%%%%%%%%%%%%%%%%%%%%%%%%%
%======================================================================================================
%
\subsection{\label{fishersZeros} Fisher's Zeros}

By introducing a discretization with an energy gap \(\varepsilon\), such that the energy of the \(k^{th}\) level can be expressed as $E_k = E_0 + k\varepsilon$, where \(E_0\) denotes the ground state energy and $k=0,1,2, \cdots$, the partition function, Eq.~(\ref{Z_total}), takes the form
\begin{equation}
Z_N(\mathcal{B})  =  e^{-\mathcal{B} E_0} \sum_{k=0}^{\Gamma} \Omega_{k} e^{-\mathcal{B} k \varepsilon},
\end{equation}
\noindent
where \(\Omega_k \equiv \Omega(E_k)\) and \(\Gamma\) is the number of energy levels. Following Fisher, we define a new variable:
\begin{equation}
x := e^{- \varepsilon \mathcal{B}} = e^{- \varepsilon \beta}  e^{- i \varepsilon \tau},
\label{x_def}
\end{equation}
\noindent
so that the partition function is now written as a polynomial,
\begin{equation}
Z = e^{- \mathcal{B} E_0} \sum_{k=0}^{\Gamma} \Omega_k x^{k} = e^{- \mathcal{B} E_0} \prod_{k=1}^{\Gamma} \left( x - x_{k} \right),
\label{polynomial}
\end{equation}
\noindent
where \(x_k\) are the zeros of the polynomial. It is worth mentioning that these roots occur in complex conjugate pairs, i.e., \(x_{k_{\pm}} = e^{- \varepsilon \beta_k} e^{\pm i \varepsilon \tau_k}\). The thermodynamic behavior of the system remains unchanged upon the introduction of a multiplicative constant to $\Omega$. Consequently, the density of states (DOS), defined as $g(E) = \Omega(E)/\sum_E \Omega(E)$, is often preferred for numerical calculations.

Since the polynomial's coefficients $\Omega_k\ge 0, \ \forall k$, any real zeros must be negative. However, it is well-known that phase transitions are defined in the thermodynamic limit. Hence, it is expected that a particular zero, or a set of zeros, will consistently approach the real positive axis as the system size increases. Those zeros are called dominant or leading zeros, and they pinch the positive real axis in the thermodynamic limit. With this fact in mind, a finite-size scaling (FSS) analysis can be employed to detect the phase transition points. The dominant zeros exhibit a power law behavior with the system size \(L\) as:
\begin{equation}
\mathrm{Im}(x_k) \propto L^{-\nu},
\label{FSS_eq}
\end{equation}
where \(\nu\) is the critical exponent of the correlation length~\cite{Itzykson}. Therefore, the analysis of the Fisher zeros consists of studying how the partition function approaches zero, i.e., $\lim_{L \to \infty} Z(\mathcal{B}_k,L) \to 0$.

%======================================================================================================
%
\subsubsection{\label{fishersZeros1order} Zeros map pattern for the first-order transition}

Recently, we have shown the connection of the unstable region of the entropy to the pattern of the Fisher zeros map~\cite{Rocha_2024}. Specifically, this region leads to a vertical line in a complex inverse temperature map. This line corresponds to a circle in the $x$-map [see Eq.~(\ref{x_def}) for the definition of $x$)]. Building upon our previous demonstration that perturbations to the linear behavior in the non-concave region of the entropy result in a negligible effect on the Fisher zero distribution, we present herein a simplified version of the demonstration detailed in Ref.~\cite{Rocha_2024}.

The demonstration here is based on the well-established double-tangent line construction across the convex region of the entropy. This construction was proposed to force $S(E)$ to obey the stability condition by eliminating the convex intruder~\footnote{For more details, see Chapter 8 in Ref.~\cite{Callen} and Section 2.7 in Ref.~\cite{Michael_book}.}. The slope of this line, $\bar{\beta}_{tan}$, can also be recognized as an estimate of the inverse transition temperature. The points of tangency define the energy range of the transition $[E', E'']$. Since heat is given by $\dbar Q = T \mathrm{d}S$, the latent heat can be given by 
\begin{equation}
    \mathcal{L} = \bar{T}_{tan}\Delta S,
    \label{latentHeat}
\end{equation}
where $\bar{T}_{tan} =1/\bar{\beta}_{tan}$ and  $\Delta S = S(E'') - S(E')$. 

Inspired by the work of Taylor {\it et al.}~\cite{Taylor}, which calculated the zeros map just in the unstable region of the entropy, we claim that
\noindent
\begin{equation}
Z'(\mathcal{B}_j) =   \sum_{E=E'}^{E''} \Omega(E)e^{-\mathcal{B}_j E} \approx 0.
\label{Z2}
\end{equation}
\noindent
This approach can be justified by the Fisher's zeros analysis that truncates the energy range~\cite{CMRocha1,Costa_2017, Mota, Ronaldo_zeros}. 

Let us consider the linear equation that describes the double-tangent line as
\begin{equation}
    S^{*}(E)  \approx S^*_{0} + \bar{\beta}_{tan}E,
\label{S_taylor}
\end{equation}
where $S^*_0$ is the value where the line intercepts the ordinate axis. The asterisk indicates that this approach for the entropy is valid only in the unstable region. In the energy range considered, $E  = E' + \ell \varepsilon$, where $\ell = 0, 1, \cdots, n'$  and $n' = (E'' - E')/\varepsilon$. Then
\begin{equation}
    S^*(E) \approx  S' + \bar{\beta}_{tan} \varepsilon \ell,
\label{Sk}
\end{equation}
\noindent
where $S' = S^*_{0} + \bar{\beta}_{tan} E'$ is a shifted constant. By inserting Eq.~(\ref{Sk}) into Eq.~(\ref{PlanckBoltzmann}), solving it for $\Omega$, then defining
\begin{equation}
 x := e^{-\varepsilon\left(\mathcal{B}-\frac{\bar{\beta}_{tan}}{k_B}\right)} = e^{-\varepsilon\left(\beta-\frac{\bar{\beta}_{tan}}{k_B}\right)} e^{-i \varepsilon \tau},
 \label{variableX}
 \end{equation}
we can rewrite Eq.~(\ref{Z2}) as
\begin{equation}
    Z' \approx e^{-\mathcal{B}F'}\sum_{\ell=0}^{n'} x^{\ell} =e^{-\mathcal{B}F'} \ \frac{1 - x^{n'+1}}{1 - x},
\label{Z2map}
\end{equation}
\noindent
where $F' = E' - S'/(k_B\mathcal{B})$. By inspecting Eqs.~(\ref{Z2map}) and (\ref{variableX}), we get $Z'= 0$ if
\begin{equation}
    \beta_{j} = \frac{\bar{\beta}_{tan}}{k_B},
\label{betaCanonico_micro}
\end{equation}
\noindent
    and
 \begin{equation}
   \tau_{j} = \frac{2\pi }{ \varepsilon (n'+1)} j\approx \frac{2\pi }{ \mathcal{L}}\ j ,
   \label{tau_j}
 \end{equation}
 \noindent
where $\mathcal{L}$ is given by Eq.~(\ref{latentHeat}) and $j = 1, 2, \cdots, n'$. The approach in the last term is valid for $n' \gg 1$ ($\varepsilon n' = E'' - E' = \Delta E$ and $\bar{T}_{tan} = \Delta E/\Delta S$). It is worth mentioning that $j \ne 0$ and $ j\ne (n'+1)$, since the denominator in the last term of Eq.~(\ref{Z2map}) requires that $x \ne 1$, hence $\mathcal{B}_{j}$ can not be a positive real number, as expected for finite systems. Furthermore, any other $j$ will lead to multiplicities and can be neglected. Since $\bar{\beta}_{tan}$ is a constant, plotting the ordered pairs ($\beta_{j}$, $ \tau_{j}$) leads to a vertical line of evenly spaced points, as previously claimed. Moreover, the distance between these zeros is inversely proportional to the latent heat. However, the coupling of $\beta$ to all Hamiltonian terms results in a less regular distribution of Fisher zeros within the complex $\beta$ plane~\cite{bena}. This inherent complexity limits the reliable calculation of zeros' distances to those located proximal to the real axis. A graphical representation of these descriptions is presented in Sec.~\ref{rc35}. 

%======================================================================================================
%
 \subsection{\label{microcanonical} Microcanonical Analysis}
The state of an isolated thermodynamic system in equilibrium is characterized by the derivatives of the entropy, eq~(\ref{PlanckBoltzmann}). As mentioned, the inverse microcanonical temperature is defined as
\begin{equation}
\bar{\beta}(\mathsf{e}) = \left( \frac{\partial \mathsf{s}}{ \partial \mathsf{e}}\right)_{\{X\}},
\label{invT_e}
\end{equation}
where $\mathsf{s}=S/N$, $\mathsf{e} = E/N$ are the entropy and the energy densities, respectively. It is worth emphasizing that we reserve the italic letter, $e$, to Euler's number.

Furthermore, the stability condition, which requires that $\mathsf{s}(\mathsf{e})$ be a monotonically increasing concave function, ensures that $\bar{\beta}$ is a monotonically decreasing, convex, and positive function. Higher-order derivatives of entropy,
\begin{eqnarray}
\hspace{-0.5cm} \gamma(\mathsf{e}) = \frac{1}{N}\left(\frac{\partial^2 \mathsf{s}}{\partial \mathsf{e}^2}\right)_{\{X\}} &
\text{ and } &
{\delta}(\mathsf{e}) = \frac{1}{N^2}\left( \frac{\partial^3  \mathsf{s}}{\partial \mathsf{e}^3}\right)_{\{X\}},
\label{highDerivatives}
\end{eqnarray}
are, respectively, an increasing concave negative function, and a decreasing convex positive function, and so on. In this work, entropy is a function of a single variable, i.e., $\mathsf{s} =  \mathsf{s}(\mathsf{e})$, the partial derivative is effectively equivalent to the total derivative and can be used interchangeably in this context. Moreover, the factor of $N^{-(k-1)}$ on the $k$-th derivatives arises from the transformation from extensive, as originally proposed, to intensive variables.

\subsubsection{The microcanonical inflection point analysis}

 Based on the criterion of minimal sensitivity~\cite{steveson_D,steveson_B}, a new method was recently proposed to characterize phase transitions by identifying least-sensitive inflection points (LSIPs) in the entropy and its derivatives~\cite{QiBachmann}.  According to this approach, in general, an independent phase transition of odd order ($2k-1$) can be identified if there is an LSIP in the ($2k-2$)-th derivative of the entropy and a corresponding minimum in the ($2k - 1$)-th derivative, i.e.,
\begin{equation}
    \left.\frac{\mathrm{d}^{2k-1}  \mathsf{s}}{\mathrm{d} \mathsf{e}^{2k-1}}\right|_{\mathsf{e}=\mathsf{e}_{tr}} > 0,
\end{equation}
where $k=1,2,\cdots$, and $\mathsf{e}_{tr}$ represents the energy at the LSIP. Notably, as mentioned in the introduction, for a first-order transition, an LSIP in the entropy results in a local minimum at $\mathsf{e} = \mathsf{e}_{tr}$ in the inverse temperature. This minimum point defines the transition temperature $\bar{T}_{tr} = 1/\bar{\beta}_{tr}$, where $\bar{\beta}_{tr}=\bar{\beta}(\mathsf{e}_{tr})$. 

Likewise, an independent phase transition of even order $2k$ occurs if there is a least-sensitive inflection point in the ($2k-1$)-th derivative of the entropy and a corresponding negative-valued maximum in the ($2k$)-th derivative, i.e.,
\begin{equation}
    \left.\frac{\mathrm{d}^{2k}  \mathsf{s}}{\mathrm{d} \mathsf{e}^{2k}}\right|_{\mathsf{e}=\mathsf{e}_{tr}} < 0.
	\label{indep_2k}
\end{equation}

Additionally, another type of transition, which occurs concomitantly with an independent transition of a lower order, can be identified. A dependent transition of even order $2k$ is signaled by the presence of an LSIP in the ($2k-1$)-th derivative of the entropy. This can be recognized by a positive-valued minimum in the ($2k$)-th derivative within the transition region of the corresponding independent transition, i.e.,
\begin{equation}
    \left.\frac{\mathrm{d}^{2k}  \mathsf{s}}{\mathrm{d} \mathsf{e}^{2k}}\right|_{\mathsf{e}=\mathsf{e}_{tr}} > 0.
\end{equation}
And a dependent transition of odd order ($2k+1$) is indicated by the presence of an LSIP in the $2k$-th derivative of the entropy and is determined by a negative-valued maximum in the ($2k+1$)-th derivative, i.e.,
\begin{equation}
    \left.\frac{\mathrm{d}^{2k+1} \mathsf{s}}{\mathrm{d} \mathsf{e}^{2k+1}}\right|_{\mathsf{e}=\mathsf{e}_{tr}} < 0.
\end{equation}
It is worth noting that the existence of an independent transition is a necessary condition for a dependent transition; however, the former can occur without the latter's presence.

Specifically, for the independent second-order phase transition, $\mathsf{e}_{tr}$ is defined at the negative-valued peak of $\gamma$. The corresponding critical temperature is then $T_C = 1/\bar{\beta}_{tr}$.  According to Behringer {\it et al.}~\cite{Behringer_2005}, the microcanonical specific heat is given by
\begin{equation}
c = -\left(\frac{\partial \mathsf{s}}{\partial \mathsf{e}}\right)^2\left(\frac{\partial^2 \mathsf{s}}{\partial \mathsf{e}^2}\right)^{-1} = -\frac{\bar{\beta}^2}{\gamma},    \label{micro_c}
\end{equation}
and the microcanonical critical exponents~\cite{Kastner2000,PhysRevE.80.051105} for specific heat $\alpha_{\varepsilon}$ and correlation length $\nu_{\varepsilon}$ are related to their canonical counterparts ($\alpha$ and $\nu$) by
\begin{equation}
\alpha_{\varepsilon} = \frac{\alpha}{1-\alpha} \ \ \text{and } \ \ \nu_{\varepsilon} = \frac{\nu}{1-\alpha}.
\label{micro_exponent}
\end{equation}
Consequently, FSS theory~\cite{Privman} predicts that the microcanonical specific heat should scale with the system size as
\begin{equation}
c \propto L^{\frac{\alpha}{\nu}}.
\label{micro_c_scale}
\end{equation}
Given this relationship and the definition of $c$ in Eq. (\ref{micro_c}),  coupled with the expectation that $\bar{\beta}$ remains finite at the transition, it follows that $\gamma$ must exhibit the following scaling behavior:
\begin{equation}
\gamma \propto L^{-\frac{\alpha}{\nu}}.
\label{micro_gamma_scale}
\end{equation}
As a consequence, in the thermodynamic limit, the value of $\gamma$ at the transition energy $e_{tr}$, denoted as $\gamma_{tr}$, approaches zero:
\begin{equation}
    \lim_{N\to \infty}{\gamma}_{tr} \to 0,
    \label{limit_gamma}
\end{equation}

In earlier studies, a positive-valued peak of $\gamma$ was also used to define a first-order phase transition~\cite{Schnabel2011,Michael_book}. Similarly, the limit presented in Eq.~(\ref{limit_gamma}) is equally valid for the earlier studies.

\subsubsection{The parametric microcanonical inflection point analysis}

A direct analogy can be drawn between Fisher zero analysis and microcanonical methods, as mathematically evident from Eqs.~(\ref{Zcondition}) and (\ref{limit_gamma}); both approaches involve investigating the behavior of a specific function as it approaches zero. Recognizing that $Z = Z(\beta)$ and ${\gamma} = {\gamma}(\mathsf{e})$, i.e., they are functions of distinct variables, we propose examining both within the framework of a unified parameter.

As alluded to in the introduction, we considered that Eq.~(\ref{invT_e}) can be solved for  $\mathsf{e}$, yielding $\mathsf{e} = \mathsf{e}({\bar{\beta}})$, which allows for the derivation of parametric curves such as $ \mathsf{s} =  \mathsf{s}({\bar{\beta}})$, ${\gamma} = {\gamma}({\bar{\beta}})$ and so forth. These parametric representations, as discussed, are not single-valued in the unstable region. Moreover, as $\bar\beta$ remains bounded at the transition point, Eq.~(\ref{limit_gamma}) continues to be valid for the parametric representation.

In this study, we propose defining a first-order transition in regions where the microcanonical parametric curves fail to exhibit the properties of a function. Specifically, for a first-order phase transition, the parametric curve for the entropy forms a $\mathsf{Z}$-like path. This behavior allows for an equal-area Maxwell construction to enforce $ \mathsf{s}({\bar{\beta}})$ being a function.

Additionally, acting in accordance with earlier microcanonical analysis studies, the analysis of ${\gamma}$ indicates a loop in the parametric curve, with the knot point serving as the indicator of the transition temperature. This loop structure effectively captures the behavior associated with the transition, complementing the insights gained from the parametric curve analysis of entropy. Moreover, the loop formation is directly associated with the undulatory curve of $\bar{\beta}(\mathsf{e})$ in the unstable region. 
This profile can be interpreted as a perturbation on the line of the Maxwell equal area construction for this curve. A decrease in the magnitude of this perturbation results in a concomitant reduction of the loop's width. 
Consequently, in the limit of zero perturbation, the loop asymptotically approaches a vertical line (an
interesting pattern that resembles the distribution of Fisher zeros in this region). It is worth mentioning that this perturbation arises from finite-size effects.

In contrast, for second-order phase transitions, the analysis of the parametric curve is consistent with conventional microcanonical analysis, characterized by a negative-valued peak in ${\gamma}({\bar{\beta}})$. The BKT transition is the most well-known example of an infinite-order phase transition.  Given the mathematical intractability of evaluating infinite derivatives, the microcanonical analysis initially appears unfeasible in this context. Nevertheless, we explored whether discernible signatures of this transition could be observed in lower-order derivatives. The following section provides graphical illustrations of all these transitions within the present framework.

%%%%%%%%%%%%%%%%%%%%%%%%%%%%%%%%%%%%%%%%%%%%%%%%%%%%%%%%%%%%%%%%%%%%%%%%%%%%%%%%%%%%%%%%%%%%%%%%%%%%%%%%%%
\section{\label{results} Results}
%%%%%%%%%%%%%%%%%%%%%%%%%%%%%%%%%%%%%%%%%%%%%%%%%%%%%%%%%%%%%%%%%%%%%%%%%%%%%%%%%%%%%%%%%%%%%%%%%%%%%%%%%%
%======================================================================================================
\subsection{\label{resultsLJ} Lennard-Jones Cluster}

We consider $N$ particles interacting via the Lennard-Jones (LJ) potential as a case study of the first-order phase transitions. The LJ potential can be written as
\begin{equation}
    U_{LJ}(r_{ij}) = 4\epsilon \left[ \left(\frac{\sigma}{r_{ij}}\right)^{12} -  \left(\frac{\sigma}{r_{ij}}\right)^6 \right],
\end{equation}
where $r_{ij} = |\mathbf{r}_j - \mathbf{r}_i|$ is the distance between the particles $i$ and $j$. We chose a reduced unit system such that $\epsilon = 1$ and $\sigma = 2^{-1/6}$; the latter was chosen to lead the minimum of the potential at distance $r_{ij} = r_0 = 1$. 

Recently, we conducted an extensive study of this model using the traditional microcanonical inflection point and the Fisher's zeros map analysis~\cite{Rocha_2024}, both derived from the DOS obtained via the replica exchange Wang-Landau method~\cite{REWL,Valentim_2015,bebe}. In this previous study, we considered $N=147$ particles confined to a sphere of radius $r_c = 4\sigma$ to reproduce the transition temperature ($T \approx 0.36$) reported in the literature~\cite{Frantsuzov}. We determined the transition temperatures to be $\bar{T}_{tr} = 0.3666(8)$ from the microcanonical inflection point, $\bar{T}_{tan}=0.364(1)$ from the double-tangent line construction, and $T_{1}/k_B=0.3622(3)$ from the leading zeros of the Fisher's zeros map. Additionally, we demonstrated the linear behavior of the dominant zeros. It is worth mentioning that, for finite systems, different quantities provide distinct transition temperatures, which converge to a single transition value as the thermodynamic limit is approached~\cite{rocha2014Procedia}. In order to authenticate the parametric approach to the microcanonical analysis, we will reuse the same set of raw data in this present study, shown in the following Section.

To illustrate the scaling behavior of $\gamma$ described by Eq.~(\ref{limit_gamma}), we simulate $N=55$ particles inside a sphere of radius $r_c=3.5\sigma$, also chosen to reproduce the transition temperature ($T \approx 0.29$) reported in the literature~\cite{Frantsuzov} Besides that, it is claimed that the results for the LJ-cluster are independent of the volume if densities ($N/V$) are  lower than that of the bulk liquid at the triple point, for the case of $147$ particles it means $r_c > 3.7\sigma$ and for $55$ particles $r_c > 2.6\sigma$~\cite{labastie}.  Therefore, to gain insight into the behavior of the Fisher's zeros pattern along the first-order transition line, we also study a system of $N=147$ particles confined within a sphere of radius $r_c = 3.5\sigma$. Both considerations are presented in Sec.~\ref{rc35}.

\subsubsection{\label{rc40} The parametric microcanonical inflection point analysis for $N=147$ particles inside a sphere of radius $r_c=4.0\sigma$}

\begin{figure}[hbt!] 
\centering
      \includegraphics[width=0.45\textwidth,keepaspectratio=true,clip]{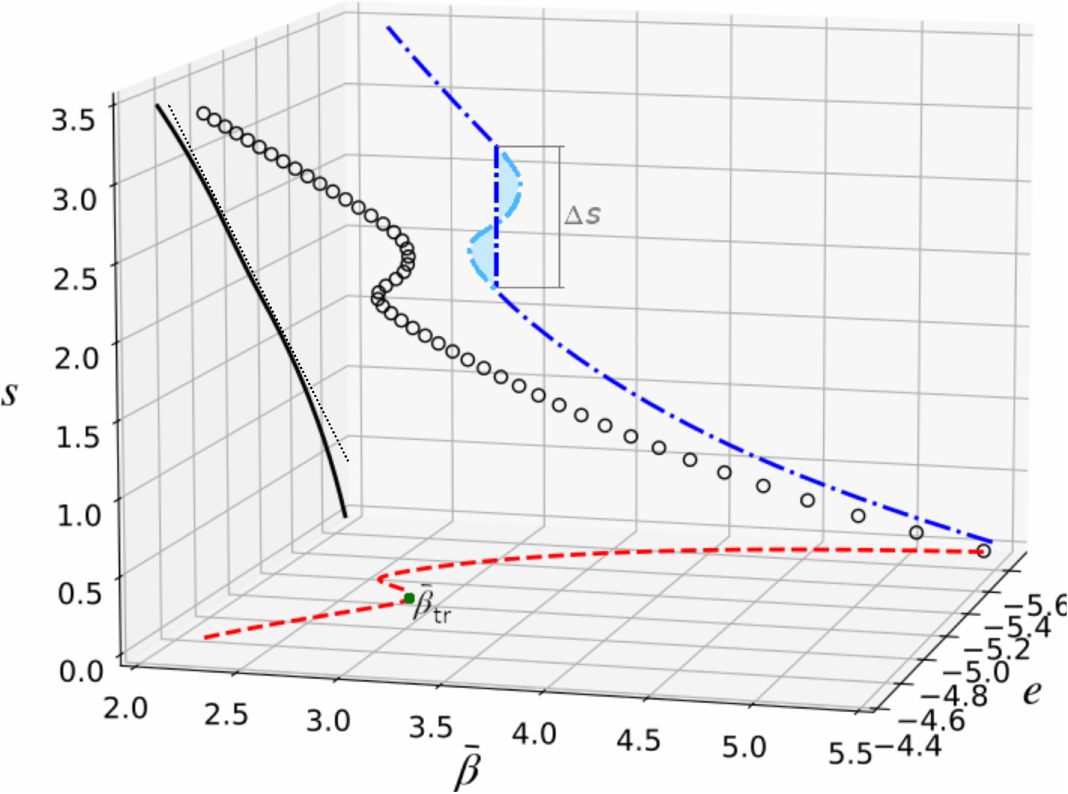}
          \caption{(Color online) Parametric curve defined by the entropy, $\mathsf{s}(\mathsf{e})$, and the microcanonical inverse temperature, $\bar{\beta}(\mathsf{e})$, for the 147-LJ cluster inside a sphere of radius $r_c=4.0\sigma$. \label{entropyXbetaXeLJ}}
\end{figure}

\begin{figure}[hbt!]
\centering
      \includegraphics[width=0.45\textwidth,keepaspectratio=true,clip]{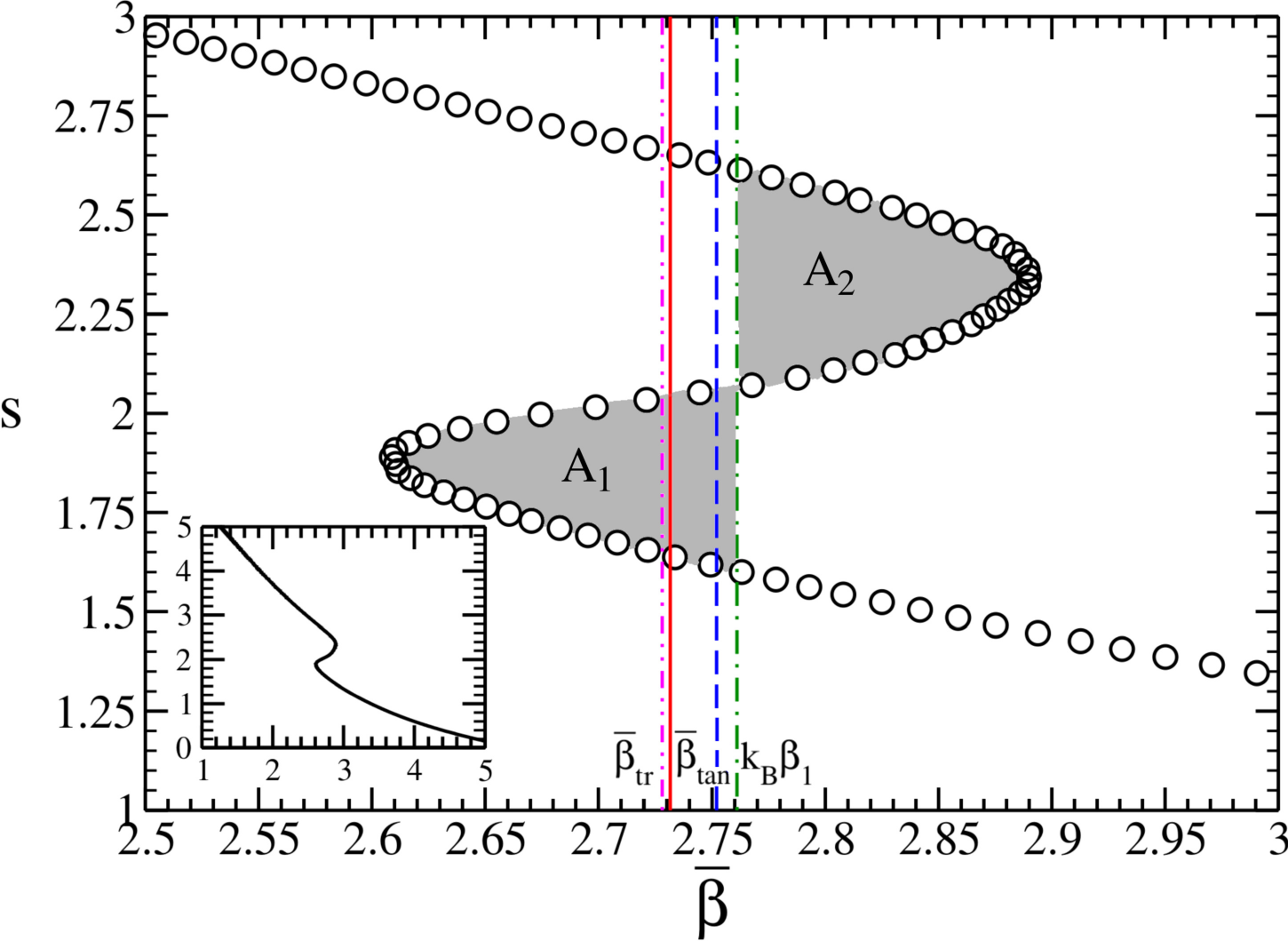}
          \caption{(Color online) Parametric curve defined by the entropy per particles, $\mathsf{s}(\mathsf{e})$, and the microcanonical inverse temperature, $\bar{\beta}(\mathsf{e})$, for the 147-LJ cluster. The inset provides a broader view of the curve. The error bars are smaller than the symbol, then not shown. The hued regions $A_1 \approx A_2$ represent the equal areas construction. The vertical lines indicate estimations of the transition temperature:  $\bar{\beta}_{tr}=2.728(6)$, shown as the double-dotted-dashed magenta line, obtained from the regular microcanonical inflection point;  $\bar{\beta}_{tan}=2.751(9)$, shown as the dashed blue line, determined by the double-tangent line construction; and $k_B\beta_1=2.761(2)$, shown as the dotted-dashed green line, identified by the Fisher´s zeros analysis. These estimations are from Ref.~\cite{Rocha_2024}. The inverse temperature at the knot position is represented by the solid red line, $\beta_{knot} = 2.732(1)$. 
        \label{entropyXbetaLJ}}
\end{figure}

To illustrate the proposal analysis, Fig.~\ref{entropyXbetaXeLJ} depicts in the solid black line the entropy per spin as a function of the energy density, $ \mathsf{s}(\mathsf{e})$, for the 147 particles inside a sphere of radius $r_c = 4.0\sigma$. It is worth emphasizing that $ \mathsf{s}(\mathsf{e})$ is the output of the REWL simulations~\footnote{The data are publicly available on the zenodo.org\cite{LJ_zenodo}}. The data presented here are the mean values of five independent simulations. Error bars were estimated via the jackknife resampling method~\cite{barkema}, and they are not shown when smaller than the symbol size. The double-tangent line construction is shown by the dotted black line. The red dashed curve in this graph represents the microcanonical inverse temperature, $\bar{\beta}(\mathsf{e})$,  obtained by the derivative of the black line, as given by Eq.~(\ref{invT_e}). For each value of $\mathsf{e}$, we plot the ordered triple $(\mathsf{e}, \bar{\beta},  \mathsf{s})$, illustrated by black circles. The projection of this curve onto the $ \mathsf{s} \times \bar{\beta}$-plane yields the parametric curve $ \mathsf{s}(\bar{\beta}),$ shown as the dotted-dashed blue curve~\footnote{The projection of the triple $(\mathsf{e}, \bar{\beta},  \mathsf{s})$ onto the $s \times \bar{\beta}$-plane is a graphical construction. We save our data into a file where the first column represents the energy density ($\mathsf{e}$), the second column represents the entropy ($ \mathsf{s}$), and the subsequent columns represent the first ($\bar{\beta}$), second ($\gamma$), and third ($\delta$) derivatives of $s$ with respect to $\mathsf{e}$. Instead of plotting $ \mathsf{s}$ against $\mathsf{e}$ we plot $ \mathsf{ \mathsf{s}}$ against $\bar{\beta}$, where $ \mathsf{ \mathsf{s}}$ and $\bar{\beta}$ correspond to the same $\mathsf{e}$. This procedure is similarly applied to the other quantities. Moreover, the curve $\mathsf{s}(\mathsf{e})$ obtained from the REWL was fitted using a Bézier curve, and the derivatives of $\mathsf{s}$ were taken as the derivatives of the Bézier curve, which are also Bézier curves, see section 4.3 in the book cited in Ref.~\cite{Michael_book}.}. This curve is detailed in Fig.~\ref{entropyXbetaLJ}, where the temperature obtained from the Fisher's zeros analysis leads to the hued regions $A_1 \approx A_2$. Therefore, the temperature of the leading zero corroborates with equal area construction, which is proposed to adjust $ \mathsf{s}(\bar{\beta})$ to comply with the uniqueness domain criteria of a function by eliminating the original points in the shaded regions and replacing them with the vertical line. Consequently, this construction leads to the discontinuity of entropy, a defining feature in the modern classification of phase transitions. A visual inspection in Fig.~\ref{entropyXbetaXeLJ} demonstrates that the equal area construction on $\mathsf{s}(\bar{\beta})$ is clearer than the double-tangent line construction on the convex intruder on $\mathsf{s}(\mathsf{e})$. This clarity facilitates the estimation of latent heat and aids in determining the order of the transition. Moreover, in Fig.~\ref{entropyXbetaLJ}, we measured the latent heat to be $\mathcal{L} = 55.1(3)$. Inserting this value into Eq.~(\ref{tau_j}), leads to $\Delta \tau = \tau_{j+1} - \tau_j = 0.1150(8)$, which differ by only $4.5\%$ from the average of the distances between the dominant zeros, $\langle \Delta \tau\rangle  = 0.110(2)$, measured on the zeros maps~\cite{Rocha_2024}. This result further supports the proposed approach.

\noindent
\noindent

\begin{figure}[hbt!] 
\centering
 \includegraphics[width=0.45\textwidth,keepaspectratio=true,clip]{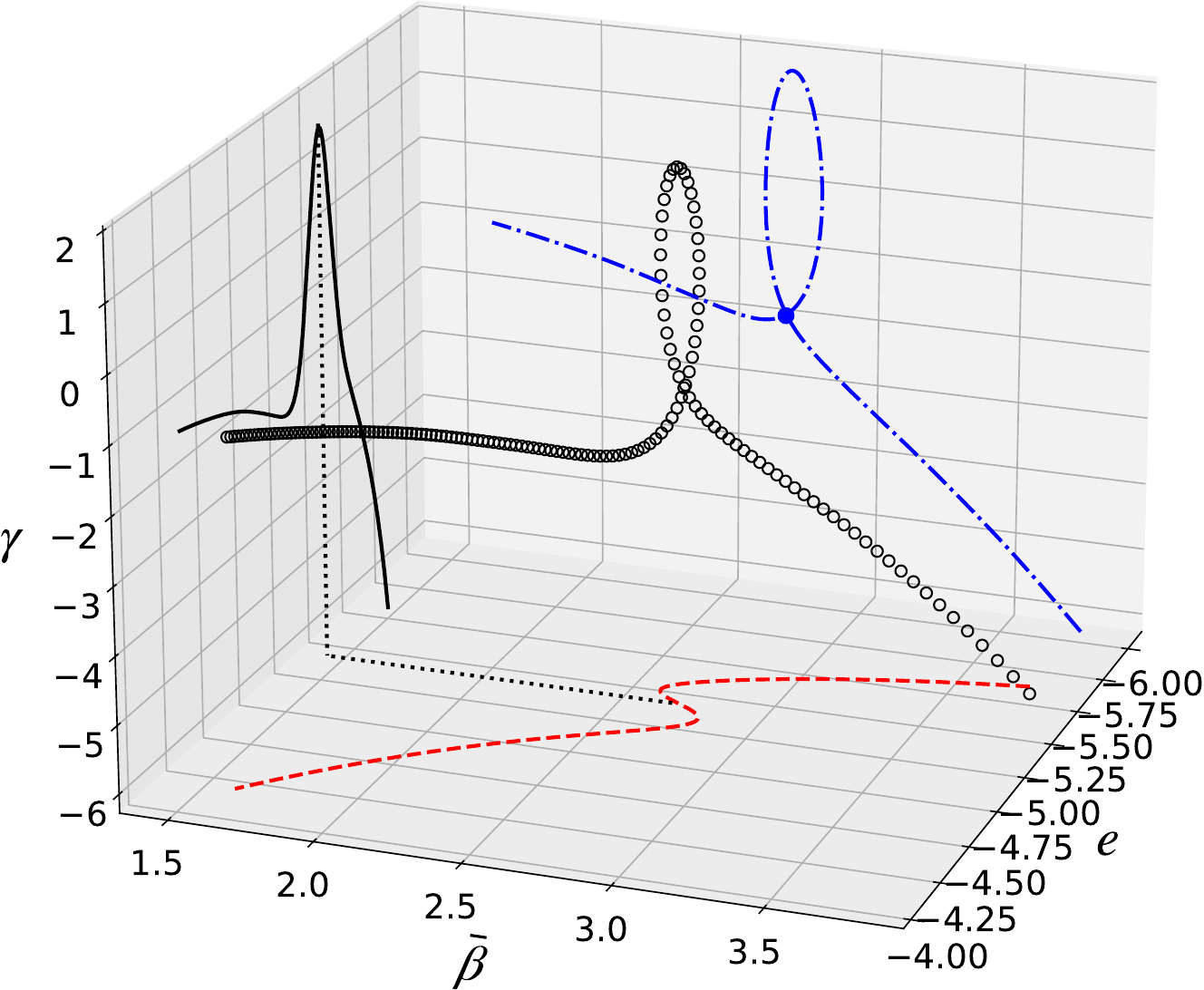}
          \caption{(Color online) Parametric curve defined by the second derivative of the entropy with respect to energy, $\gamma(\mathsf{e})$, and the microcanonical inverse temperature, $\bar{\beta}(\mathsf{e})$, for the 147-LJ cluster. \label{gammaXbetaXeLJ}}
\end{figure}

\begin{figure}[hbt!] 
\centering
      \includegraphics[width=0.45\textwidth,keepaspectratio=true,clip]{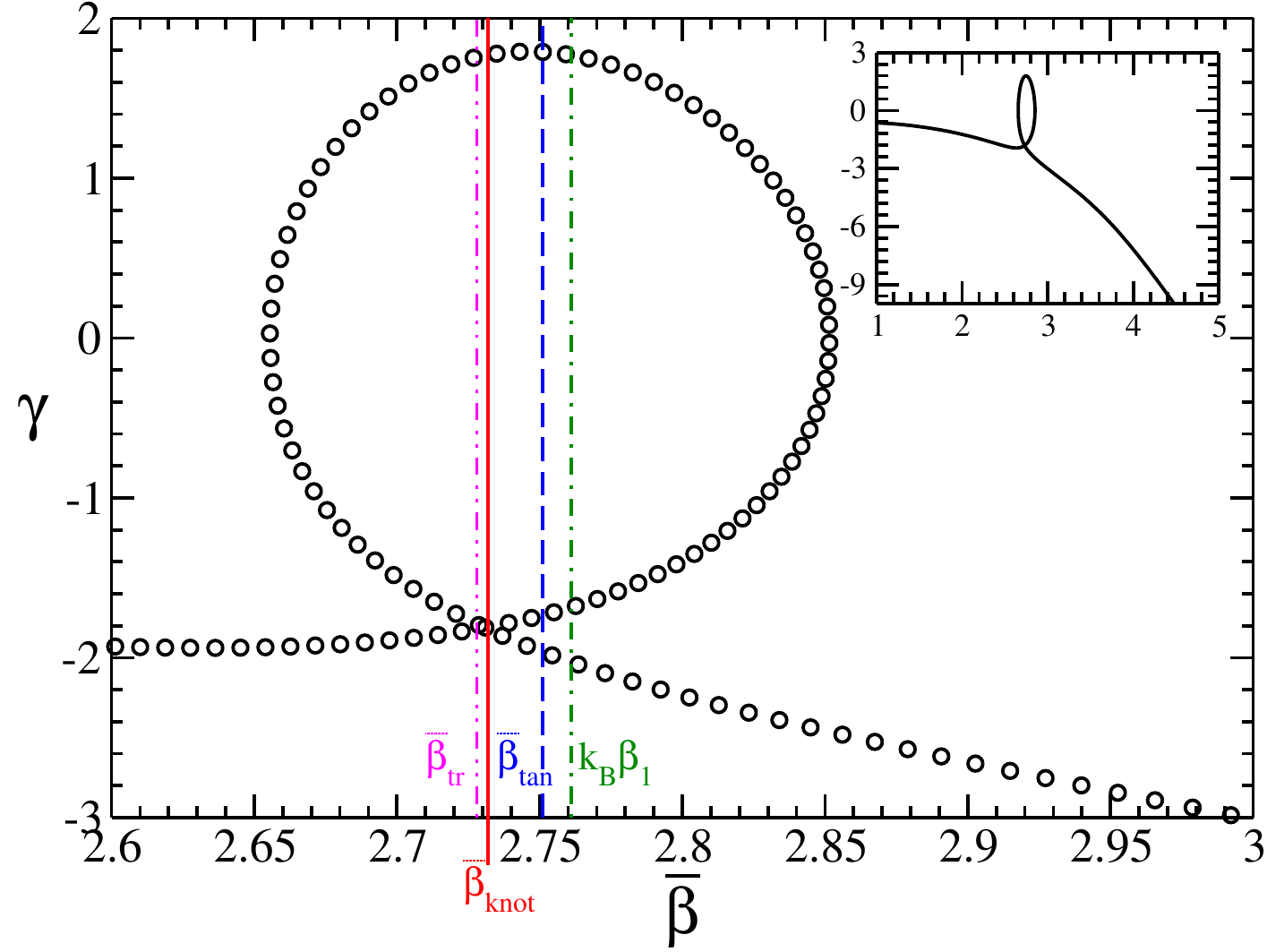}
          \caption{(Color online) Parametric curve defined by the second derivative of the entropy with respect to energy, $\gamma(\mathsf{e})$, and the microcanonical inverse temperature, $\bar{\beta}(\mathsf{e})$, for the 147-LJ cluster. The inset provides a broader view of the curve. Error bars smaller than the symbol size are not shown. The vertical lines scheme is shown in the caption of Fig.~\ref{entropyXbetaLJ}. \label{gammaXbetaLJ}}
\end{figure}

Figure~\ref{gammaXbetaXeLJ} demonstrates a similar process for obtaining the parametric curve $\gamma(\bar{\beta})$, which is shown in detail in Fig.~\ref{gammaXbetaLJ}. Additionally, Fig.~\ref{gammaXbetaXeLJ} illustrates the regular microcanonical inflection point analysis by the dotted black line. This line projects the energy of the peak position of $\gamma(\mathsf{e})$ onto $\bar{\beta}(\mathsf{e})$, leading to $\bar{\beta}_{tr} = \bar{\beta}(\mathsf{e}_{tr})$. The loop on the parametric curve $\gamma(\bar{\beta})$ for the first-order transition is illustrated in this figure. 
By applying the uniqueness domain criterion for functions, the curve is truncated at the knot position, thereby eliminating the loop points and defining the transition temperature at this position. In  Fig.~\ref{gammaXbetaLJ}, we measured the temperature of the knot position to be $\bar{T}_{knot}=0.3660(1)$. It is worth mentioning that $|\bar{T}_{tr} - \bar{T}_{knot}|$ is smaller than the error of $\bar{T}_{tr}$, where $\bar{T}_{tr}$ is the transition temperature obtained from the regular microcanonical analysis. Additionally, the latent heat was calculated using all estimated transition temperatures, with the results differing within the error bars. Specifically, in addition to the previously mentioned $\mathcal{L}_1 = 55.1(3)$ obtained from $\beta_1$, we find that $\beta_{tr}$ yields $\mathcal{L}_{tr} =55.0(6)$, $\beta_{tan}$ estimates $\mathcal{L}_{tan} =54.2(7)$, and $\beta_{knot}$ measures $\mathcal{L}_{knot} =54.7(2)$. This yields an average value of $\mathcal{L}_{avg} =54.8(5)$.

\noindent

\subsubsection{\label{rc35} Parametric microcanonical inflection point and Fisher's zeros analysis for $N=55$ and $N=147$, inside a sphere of radius $r_c = 3.5\sigma$}

\begin{figure}[hbt!] 
\centering \includegraphics[width=0.45\textwidth,keepaspectratio=true,clip]{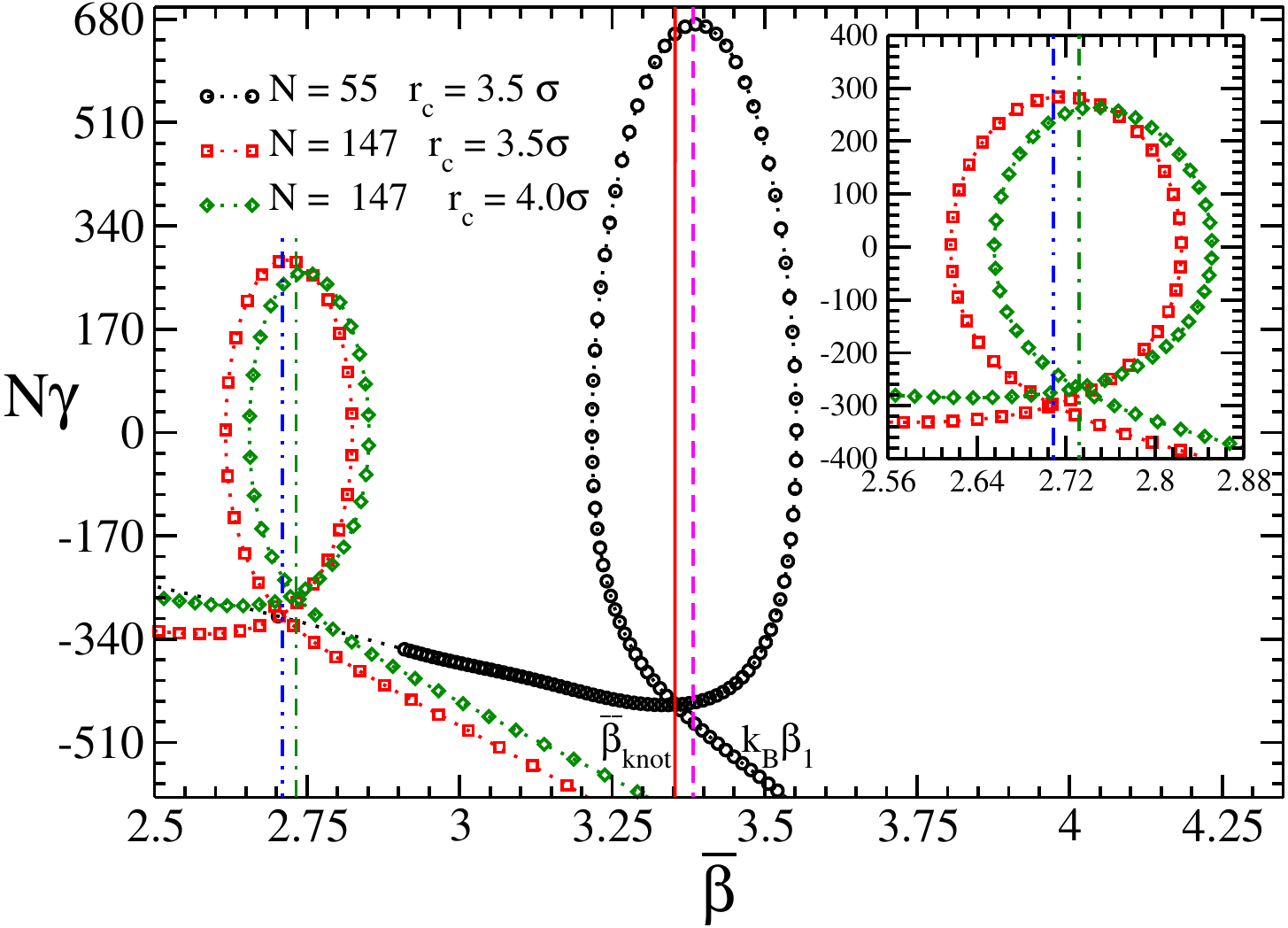}
          \caption{(Color online) Illustration of the scale behavior of $\gamma$ described by Eq.~(\ref{limit_gamma}). The inset shows a zoom in the loop of the results for $N=147$ particles. \label{gammaXbeta_FSS}}
\end{figure}

Figure~\ref{gammaXbeta_FSS} depicts the scaling behavior of $\gamma(\bar{\beta})$ with system size. The data for $L=55$ particles confined within a spherical volume of radius $r_c=3.5\sigma$ are represented by black circles. The knot position corresponds to the transition temperature, $\bar{T}_{knot} = 0.2982(2)$, as indicated by the solid red line. For comparison, the transition temperature associated with dominant zeros is shown by the dashed magenta line. Results for $N=147$ particles are displayed as red squares for $r_c=3.5\sigma$ and green diamonds for $r_c=4.0\sigma$. 

A clear shrinkage of the loop toward $\gamma=0$ is observed with increasing system size, consistent with the expectation from Eq.~(\ref{limit_gamma}). Notably, the loop is slightly smaller for $L=147$ at $r_c=4.0\sigma$ compared to $r_c=3.5\sigma$, as evident in the inset of the figure. It is well established that for simple systems, the first-order transition line tends toward a critical point with increasing pressure and, consequently, a decrease in latent heat. A systematic investigation of this aspect falls outside the scope of the present study. Primary results on highly compressed configurations have exhibited complex behavior characterized by two distinct transition signals, resembling observations for LJ clusters with an intermediate number of particles between the established ``magic numbers" (the chosen values of $N=55$ and $N=147$ are recognized as magic numbers). While LJ clusters corresponding to magic numbers typically display a strong first-order transition from the liquid phase to the energetically preferred low-temperature Mackay icosahedral solid package,  intermediate-sized clusters display an additional solid-solid transition. This interjacent transition occurs between anti-Mackay (hexagonal close-packed, HCP, overlayers) and Mackay (face-centered cubic, FCC, overlayers) icosahedral shell packaging~\cite{northby}.

\begin{figure}[hbt!] 
\centering
\begin{tabular}c
\includegraphics[width=0.45\textwidth,keepaspectratio=true,clip]{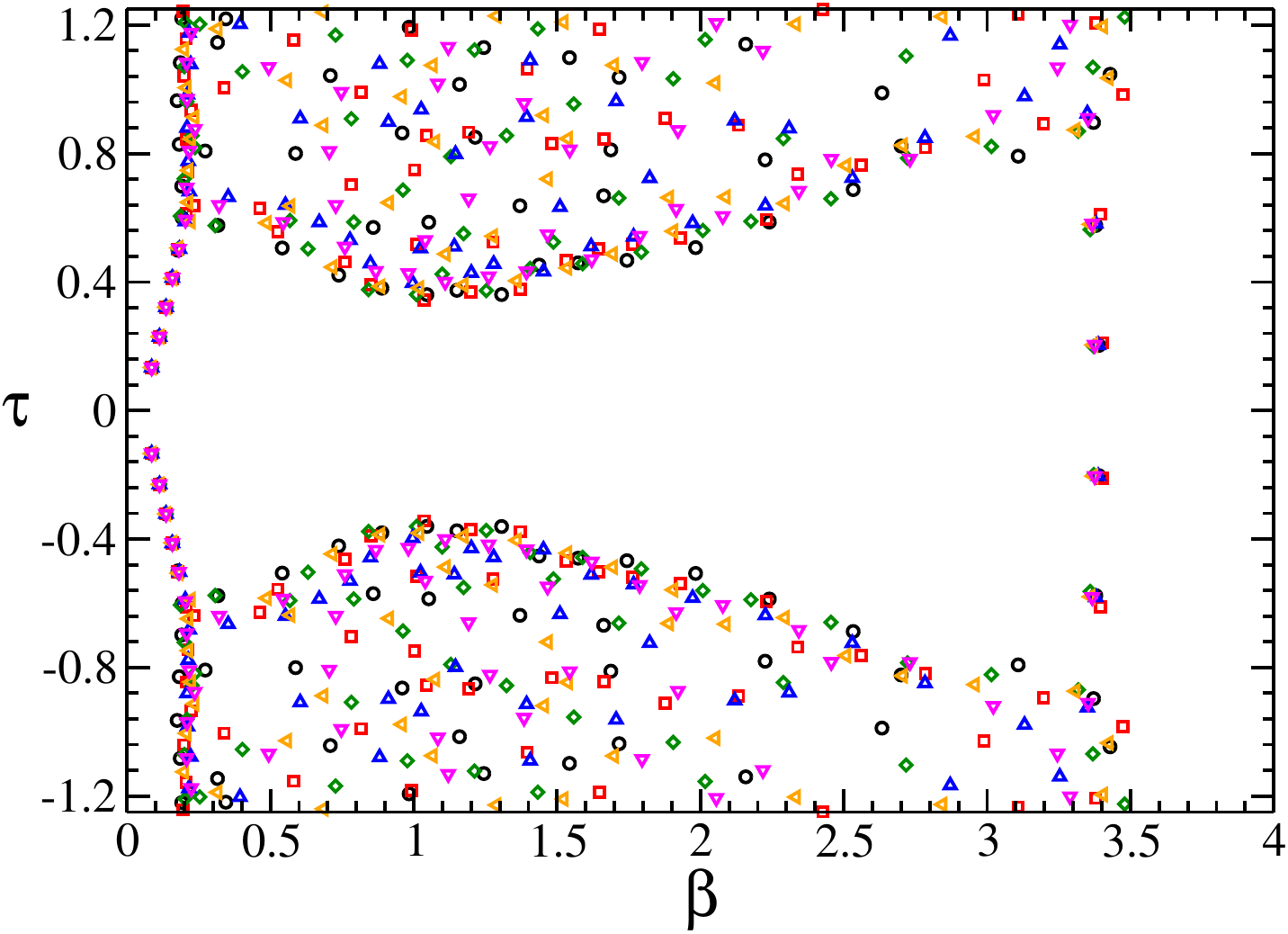}       \\ (a) \\
\includegraphics[width=0.45\textwidth,keepaspectratio=true,clip]{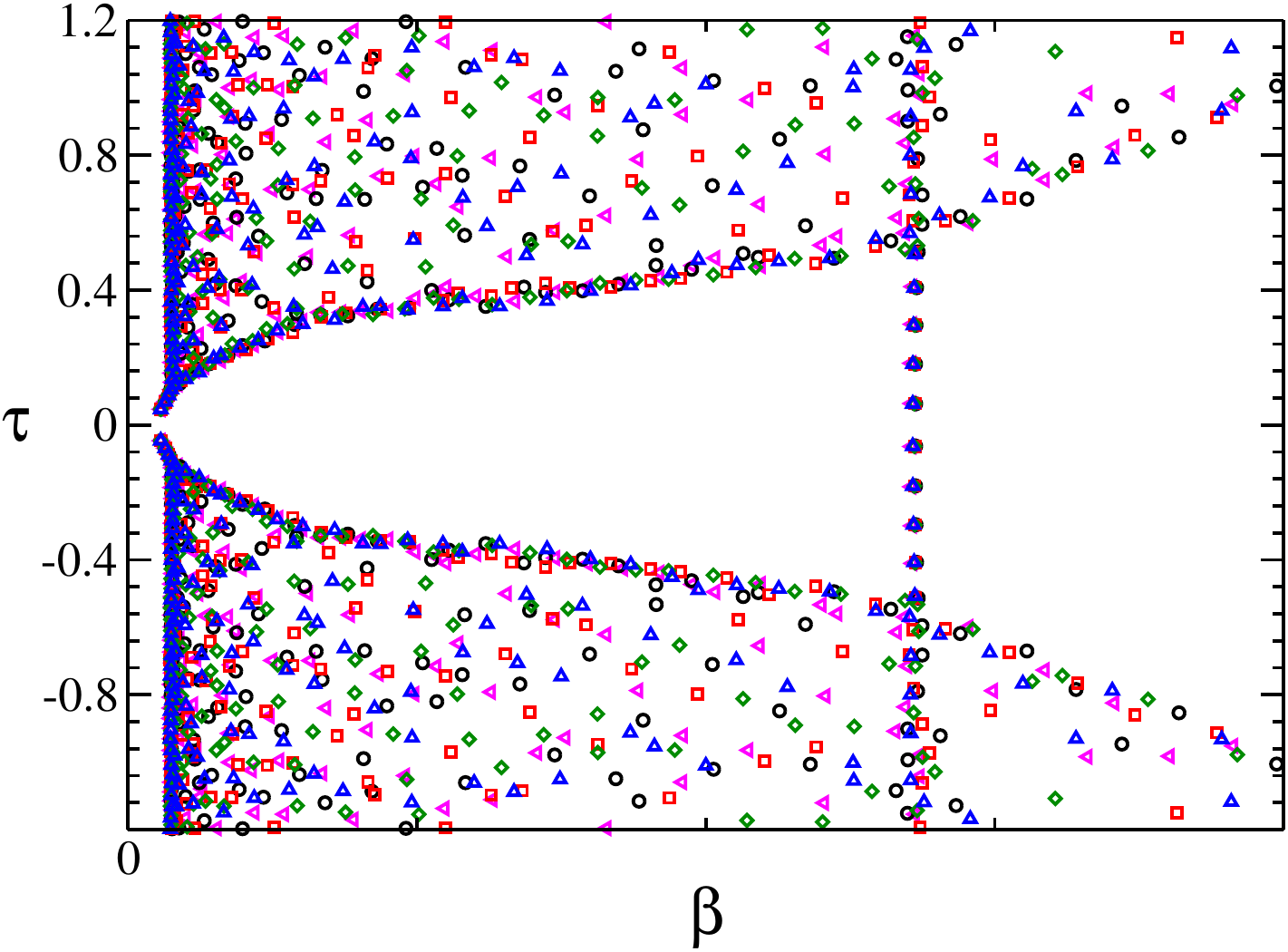} \\ (b) 
\end{tabular}
          \caption{(Color online) Fisher's zeros map for the LJ cluster constraint to a sphere of radius $r_c=3.5\sigma$. In  panel (a) for $N=55$ particles and in panel (b) for $N=147$ particles  (see Ref~\cite{Rocha_2024} for 147-LJ cluster in $r_c=4.0\sigma$). Each symbol corresponds to a map obtained from an independent simulation.\label{zerosLJ}}
\end{figure}

Figure~\ref{zerosLJ} presents the Fisher's zeros map for the Lennard-Jones (LJ) cluster constrained to a sphere of radius $r_c = 3.5\sigma$. We use MPSolve~\cite{mpsolve1,mpsolve2} as the root finder. A characteristic pattern of equally spaced vertical lines of zeros is observed, converging toward the real temperature axis, indicative of a first-order phase transition. Panel (a) displays the map for $N=55$ particles, where the real part of the leading zero corresponds to a transition temperature of $k_BT_1 = 0.2956(4)$ and latent heat of $\mathcal{L} = 30.80(3)$. Panel (b) shows the map for $N=147$ particles, with a measured transition temperature of $k_BT_1 = 0,3676(4)$ and latent heat of $\mathcal{L} =51.51(3)$. The zeros map for a 147-particle LJ cluster confined within a sphere of radius $r_c=4.0\sigma$ can be found in reference~\cite{Rocha_2024}. Each symbol in the figure represents a map obtained from an independent simulation. It can be noted that, as it is well-known, the zeros of the partition function are highly sensitive to statistical fluctuations, except for the dominant zeros.

\subsection{\label{resultsIsing} 2D Ising Model}

The two-dimensional (2D) Lenz-Ising model serves as a theoretical framework for the second-order phase transitions~\cite{brush}. This model considers spin-$1/2$ particles, arranged in a fixed lattice, that interact with their nearest neighbors. Lars Onsager derived the exact solution for the square lattice in 1944~\cite{Onsager}, where the critical temperature is deduced to be  $T_c = 2/\ln(1 + \sqrt{2})$.

For the case where the external magnetic field is zero, the Hamiltonian describing the model is given by
 \begin{equation}
\mathcal{H} = -J\sum_{\langle i,j \rangle} \sigma_i\sigma_j,
\label{hamiltonian_ising}
\end{equation}
where $J$ is the exchange integral (positive for ferromagnetic and negative for antiferromagnetic interactions), and ${\sigma}_i = \pm |\sigma|$ represents the spin at site $i$. 
Throughout this section, energy is expressed in units  $J|\sigma|^2$ and the canonical temperature in units of $J|\sigma|^2/k_B$. We consider a $L\times L$ square lattice with periodic boundary conditions, where $L$ is the linear lattice size. The notation $\langle i,j \rangle$ indicates summation over nearest-neighbor pairs. 

\subsubsection{\label{pMIP_ising} Parametric microcanonical inflection point analysis for the 2D Ising model}

Recent studies employing conventional microcanonical inflection point analysis of this model~\cite{QiBachmann,Sitarachu_2020,Sitarachu_2020B,Sitarachu_2022} have reported evidence of higher-order phase transitions in addition to the well-known second-order ferromagnetic/paramagnetic transition. Specifically, two additional transitions were identified: a dependent transition occurring above the critical temperature and an independent transition occurring below it, as illustrated in the following paragraphs.

\begin{figure}[hbt!]
\centering
\includegraphics[width=0.45\textwidth,keepaspectratio=true,clip]{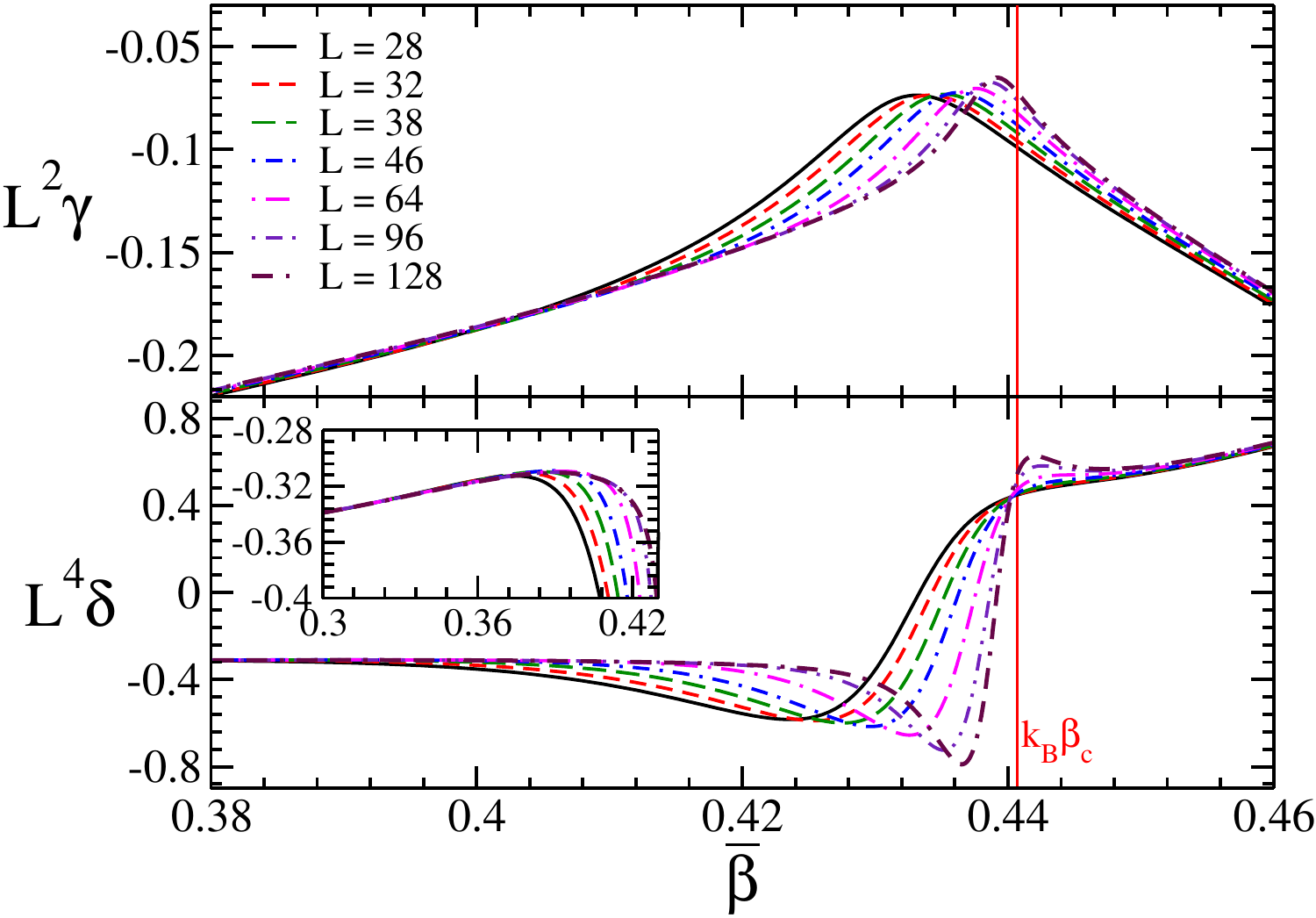} 
    \caption{(Color online) $\gamma(\bar{\beta})$ and $\delta(\bar{\beta})$ for the 2D Ising model on the square lattice. The top panel presents $N\gamma(\bar{\beta})$ and the bottom presents $N^2\delta(\bar{\beta})$) for systems sizes ranging from $L=28$ up to $L=128$. The red solid vertical line indicates the critical inverse temperature $\beta_c = \ln{(1 + \sqrt{2})}/2$. \label{MicroIsing96} }
\end{figure}

Figure~\ref{MicroIsing96} displays the results of a parametric microcanonical inflection point analysis applied to the two-dimensional Ising model, for linear system sizes ranging from $L=28$ to $128$. The DOS used in this analysis was obtained from Beale's exact solution \cite{Beale}.

\begin{figure}[hbt!]
\centering
\includegraphics[width=0.475\textwidth,keepaspectratio=true,clip]{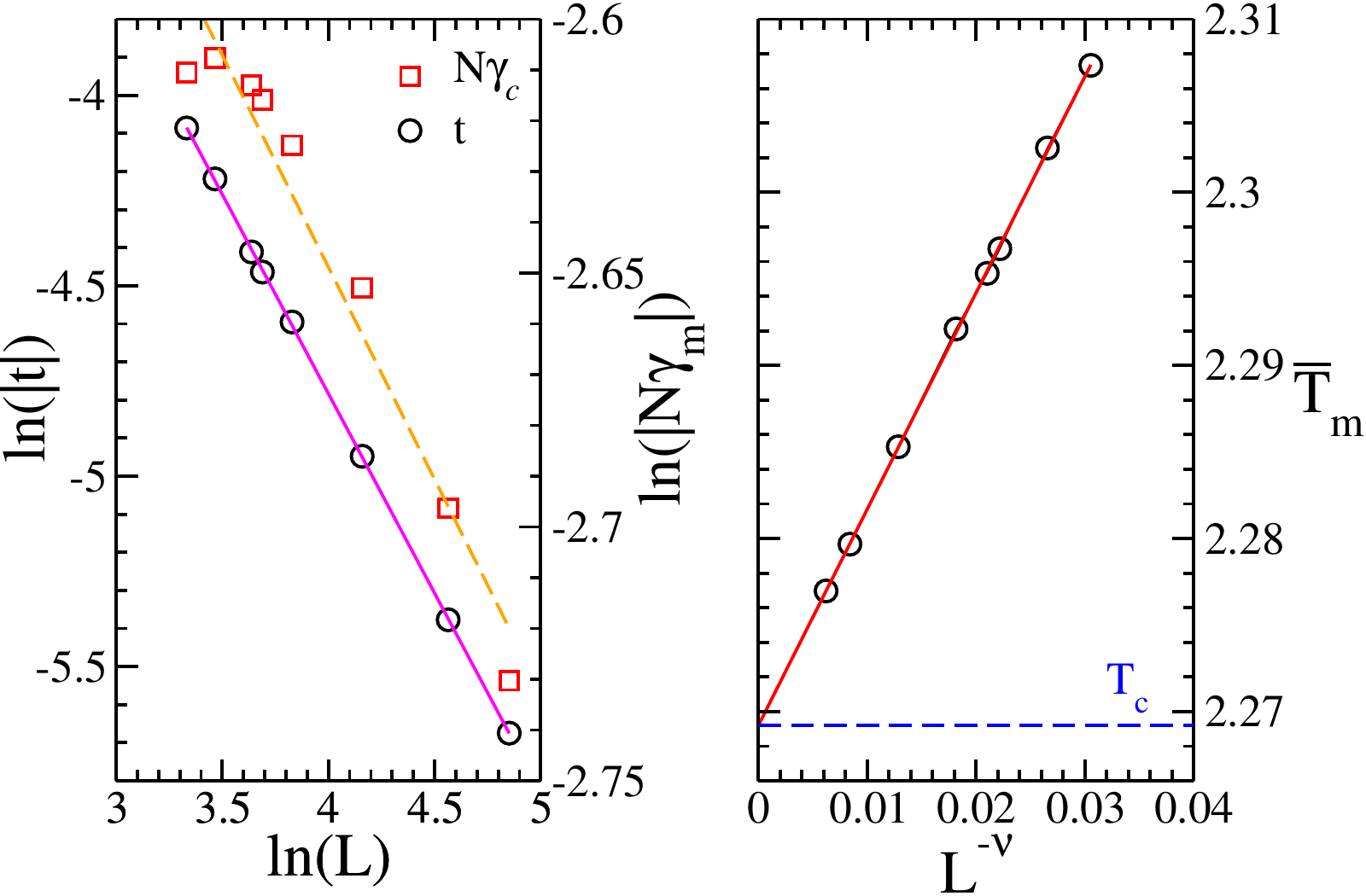} 
    \caption{(Color online)  Fisher's Zeros map for the Ising model with linear system size $L=96$. The red dashed squares indicate the leading zeros.
        \label{fss_ising}}
\end{figure}

The top panel of Fig.~\ref{MicroIsing96} illustrates the behavior of $N\gamma(\bar{\beta})$. A FSS analysis for the peak position of $\gamma$, denoted as $\gamma_m$, is presented in Fig.~\ref{fss_ising}. The left panel of the figure shows both $\ln(|N\gamma_m|)$ versus $\ln(L)$ and $\ln(|t|)$ versus $\ln(L)$, where $t=1-T_m/T_c$ is the reduced temperature at the peak position, $T_m$. Comparing the former plot with Eq.~(\ref{micro_gamma_scale}), the slope of the linear regression yields a critical exponent $\alpha/\nu = 0.083(8)$. This value aligns with the 2D Ising universality class, for which $\alpha = 0$. Given that $t \propto L^{-\nu}$, the linear regression of the latter plot provides $\nu = 1.047(4)$, which is close to the predicted value of $\nu=1$ for the 2D Ising universality class.

The right panel of Fig.~\ref{fss_ising} plots $T_m$ against $L^{-\nu}$, utilizing the previously estimated $\nu$ value. The critical inverse temperature extrapolated from this linear regression is $T_c = 2.2692(1)$. This value exhibits a deviation of approximately $6.5 \times 10^{-4}\%$ from the exact critical value, $T_c \approx 2.269185$, which is marked by the horizontal blue dashed line.  It is known that deviations from the simple scaling form for small system sizes~\cite{landau_1976} necessitate logarithm corrections to properly describe the criticality in this limit~\cite{johnston2006a, johnston2006b, moueddene,e26030221}. However, the estimates for critical exponents and the high accuracy in determining the transition temperature presented here corroborate the reliability of the analysis.

 \noindent
\noindent

The bottom panel of Fig.~\ref{MicroIsing96} illustrates the behavior of $N^2\delta(\bar{\beta})$. For system sizes $L=96$ and $128$, we observe positive-valued local minima in $\delta$ within the region where $\bar{\beta} > \beta_c$. The appearance of these minima suggests an independent third-order phase transition. This finding aligns with previous research indicating that a fourth-order transition, defined by Eq.~(\ref{indep_2k}), observed in smaller systems, evolves into this third-order transition as the system size $L$ increases \cite{QiBachmann}. Due to limitations in the availability of the DOS for larger system sizes in this study, an FSS analysis could not be performed for this particular transition.

Drawing parallels with the Binder-cumulant analysis~\cite{Binder1981} and the established use of Lee-Yang zero ratios across different spatial volumes~\cite{wada}, the critical point can be estimated by the intersection of $\delta(\bar{\beta})$ at zero.
Furthermore, the inset in the bottom panel of Fig.~\ref{MicroIsing96}(b) reveals a negative peak across all system sizes investigated. This feature corresponds to an additional dependent third-order transition within the paramagnetic phase. An attempt at FSS analysis for this transition did not exhibit any scaling behavior, consistent with previous microcanonical studies~\cite{QiBachmann}, and thus these results are not presented.

\noindent
\noindent

\subsubsection{\label{pMIP_ising_critical} Fisher leading zero critical behavior analysis}

\begin{figure}[hbt!]
\centering
\includegraphics[width=0.475\textwidth,keepaspectratio=true,clip]{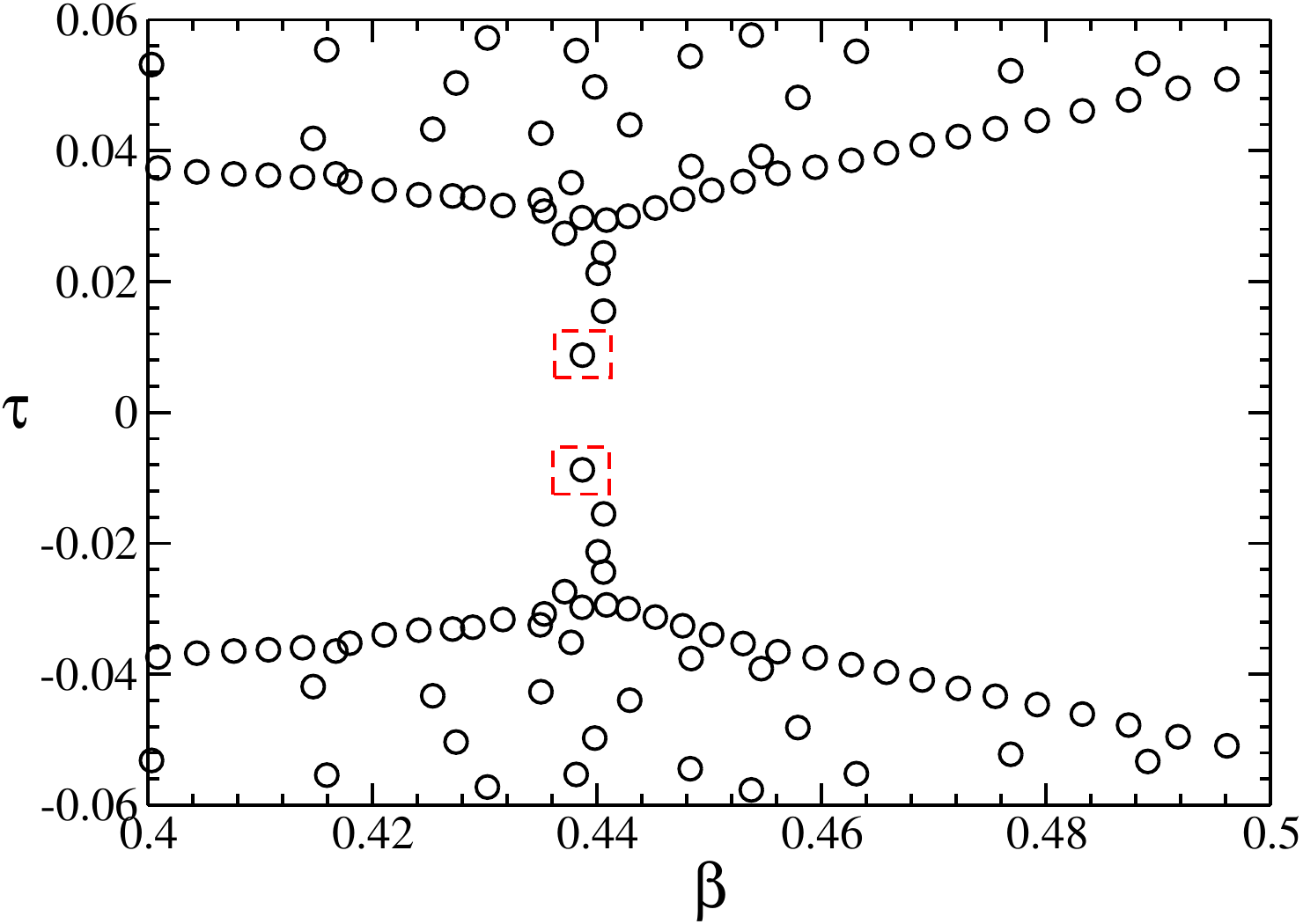} 
    \caption{(Color online)  Fisher's zeros map for the Ising model with linear system size $L=96$. The red dashed squares indicate the leading zeros.
        \label{ZerosIsing96}}
\end{figure}

Figure~\ref{ZerosIsing96} depicts the Fisher's zeros map for the Ising model on the complex $\mathcal{B}$-plane, an alternative representation to the conventional $x$-complex plane~\cite{drugowich} or analogous quantities such as $w = 2\sinh{(2\beta)}$~\cite{johnston}. Notably, circles in the $x$-complex plane are mapped onto vertical lines in the $\mathcal{B}$-plane. Furthermore, circles with radii exceeding unity correspond to negative temperatures ($\beta < 0$), which are not displayed in our graphs. For this model, negative temperatures mean antiferromagnetic ground states, as the temperature is measured in units of $J|\sigma|^2/k_B$. Owing to the symmetry of DOS for the Ising model, the magnitudes of the transition temperatures are identical. In Fig.~\ref{ZerosIsing96}, the real part of the leading zero is determined to be $k_B{\beta}_{1} \approx 0.43868$. While exhibiting a vertical line pattern, this map deviates from the characteristic pattern observed in first-order transitions due to the non-uniform spacing of the dominant zeros~\cite{Borrmann, nelson}. 

From the FSS analysis, guided by Eq.~(\ref{FSS_eq}), the imaginary component of the zeros is observed to approach the real axis, with an extrapolated value $\tau_c=1.6(5)\times10^{-4}$.  Furthermore, this analysis yields a value of $\nu=0.993$ for the critical exponent. This result is in agreement with the expected value of $\nu=1$. Moueddene {\it et al.} recently indicated that critical coefficients derived from the zeros of the partition function 
can effectively predict critical phenomena while maintaining a reasonably low computational cost~\cite{Moueddene_2024}. In our analysis, the critical inverse temperature estimated from the leading zeros is $\beta_c = 0.440589(2)$, which deviates by approximately $0.02\%$ from the exact value - thus corroborating the previous analysis.
\noindent
\noindent

%======================================================================================================
\subsection{\label{resultsXY} XY Model}

The Berezinskii-Kosterlitz-Thouless (BKT) transition, exemplified by the XY model, is a topological phase transition driven by the unbinding of vortices at temperature $T_{BKT}$. It is characterized by the absence of discontinuities or divergences in any finite-order derivative of the free energy.  The XY model describes two-dimensional systems, a prototype being a lattice of spins with continuous symmetry. The state of each spin is characterized by an angular variable, $\theta_i$, denoting its orientation with respect to a fixed reference axis within the plane. Interactions are confined to nearest neighbors, and the Hamiltonian of the system assumes the following form:
\begin{equation}
    \mathcal{H} = -J\sum_{\langle i, j\rangle} \cos{(\theta_i - \theta_j)}.
\end{equation}

\begin{figure}[hbt!]
\centering
\includegraphics[width=0.475\textwidth,keepaspectratio=true,clip]{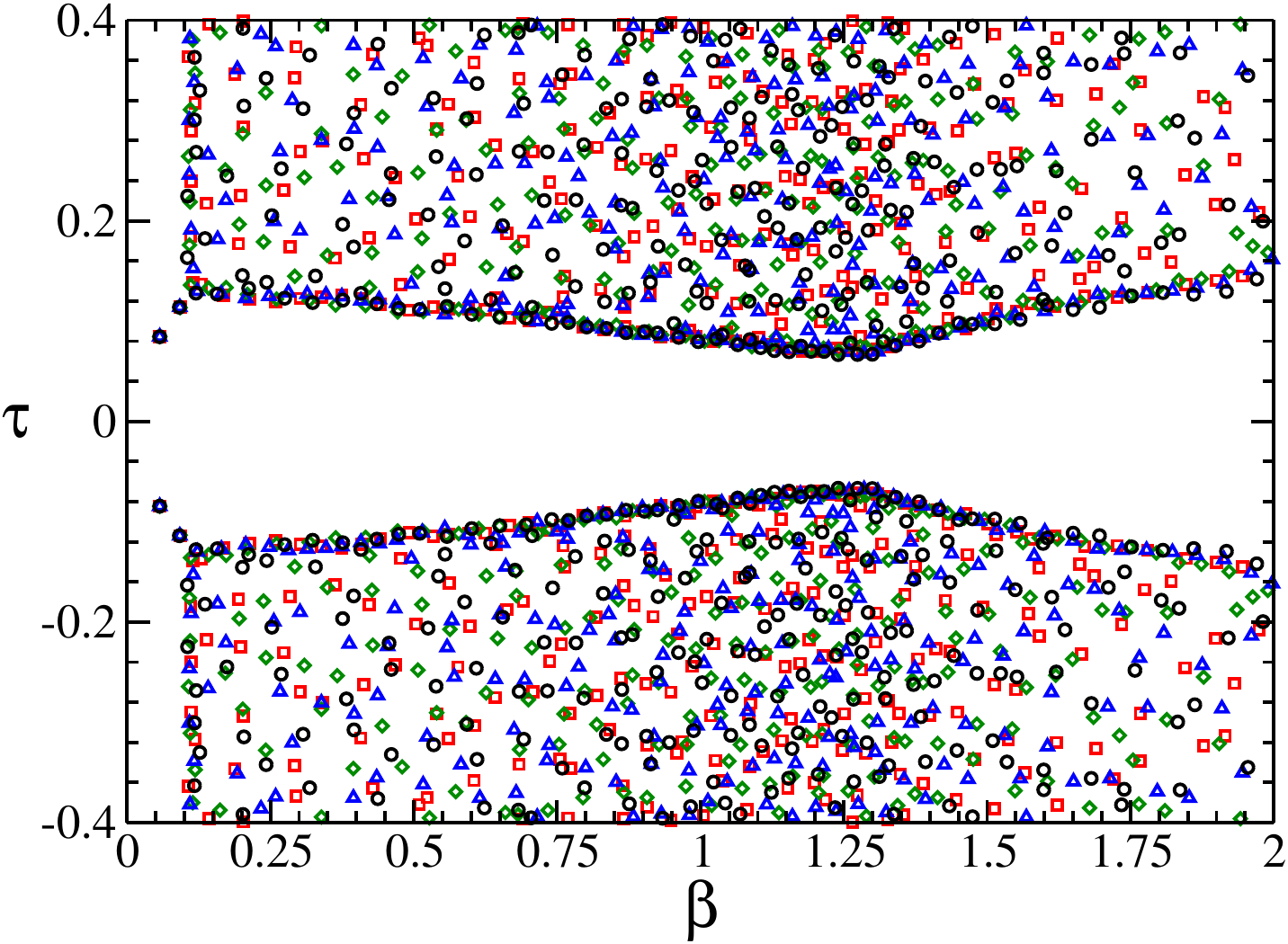} 
    \caption{(Color online)  Fisher's zeros map for the XY model with linear system size $L=50$. Each symbol corresponds to a map obtained from an independent simulation.
        \label{ZerosXY60}}
\end{figure}

In previous work, we investigated this model on a square lattice of dimension $L\times L$ employing Fisher's zeros methodology~\cite{RMCosta,tati,Costa2019}. An example of these mappings on the \(\mathcal{B} \) plane, instead of the originally complex-$x$ map, is depicted in Fig.~\ref{ZerosXY60}. Despite the absence of dominant zeros, the BKT transition was identified by examining the internal border of zeros through FSS analysis, where the BKT transition temperature was estimated to be $T_{BKT}=0.704(3)$. Our findings were in complete concordance with theoretical predictions, enabling the classification of the phase transition as belonging to the BKT universality class~\footnote{The raw data ($S\times E$), used in this study, are publicly available on zenodo.org~\cite{XY_zenodo}}. 

\noindent
\noindent

\begin{figure}[hbt!]
\centering
\begin{tabular}{c}
\includegraphics[width=0.45\textwidth,keepaspectratio=true,clip]{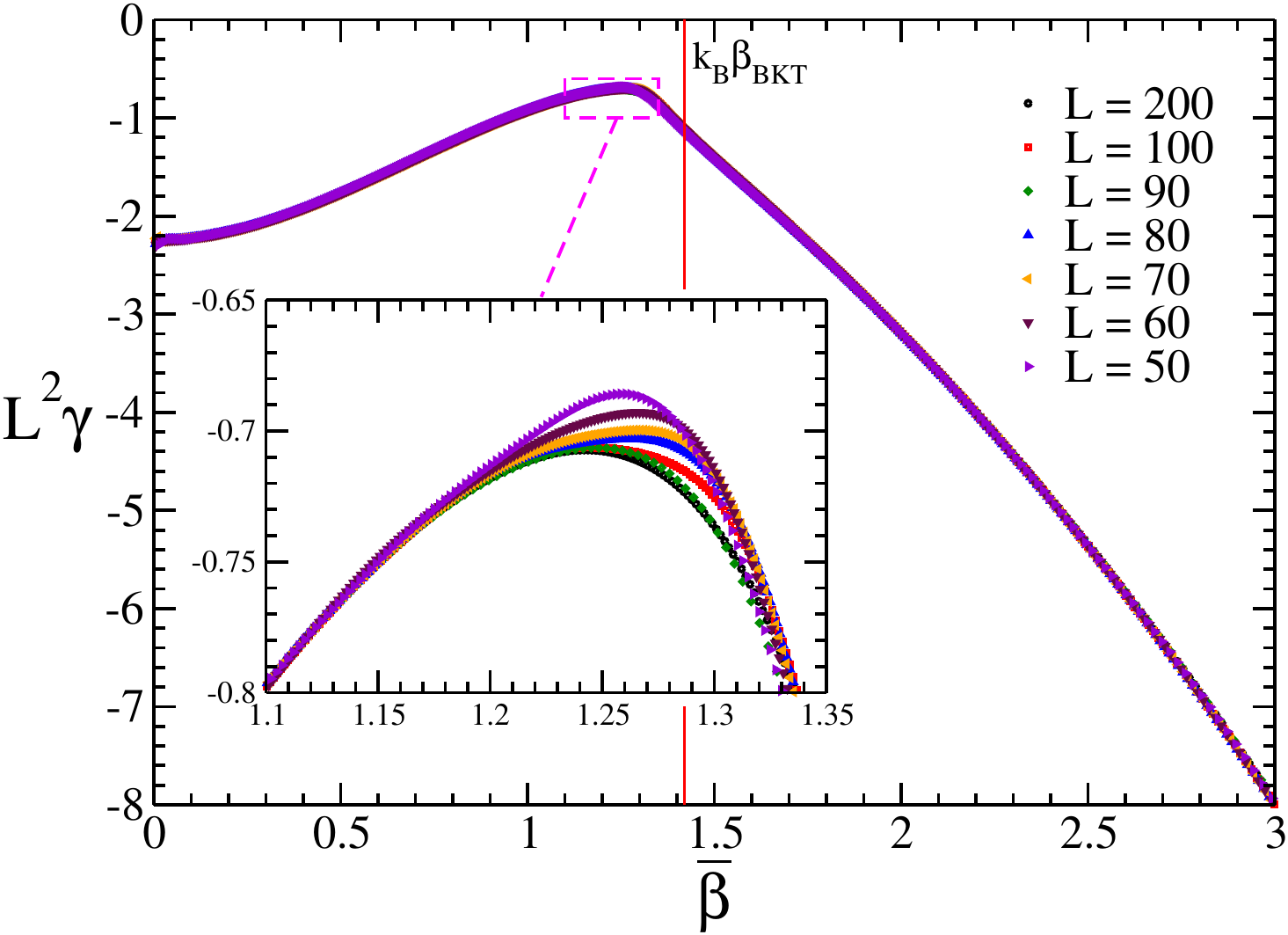} \\ (a)  \\  \\ 
\includegraphics[width=0.45\textwidth,keepaspectratio=true,clip]{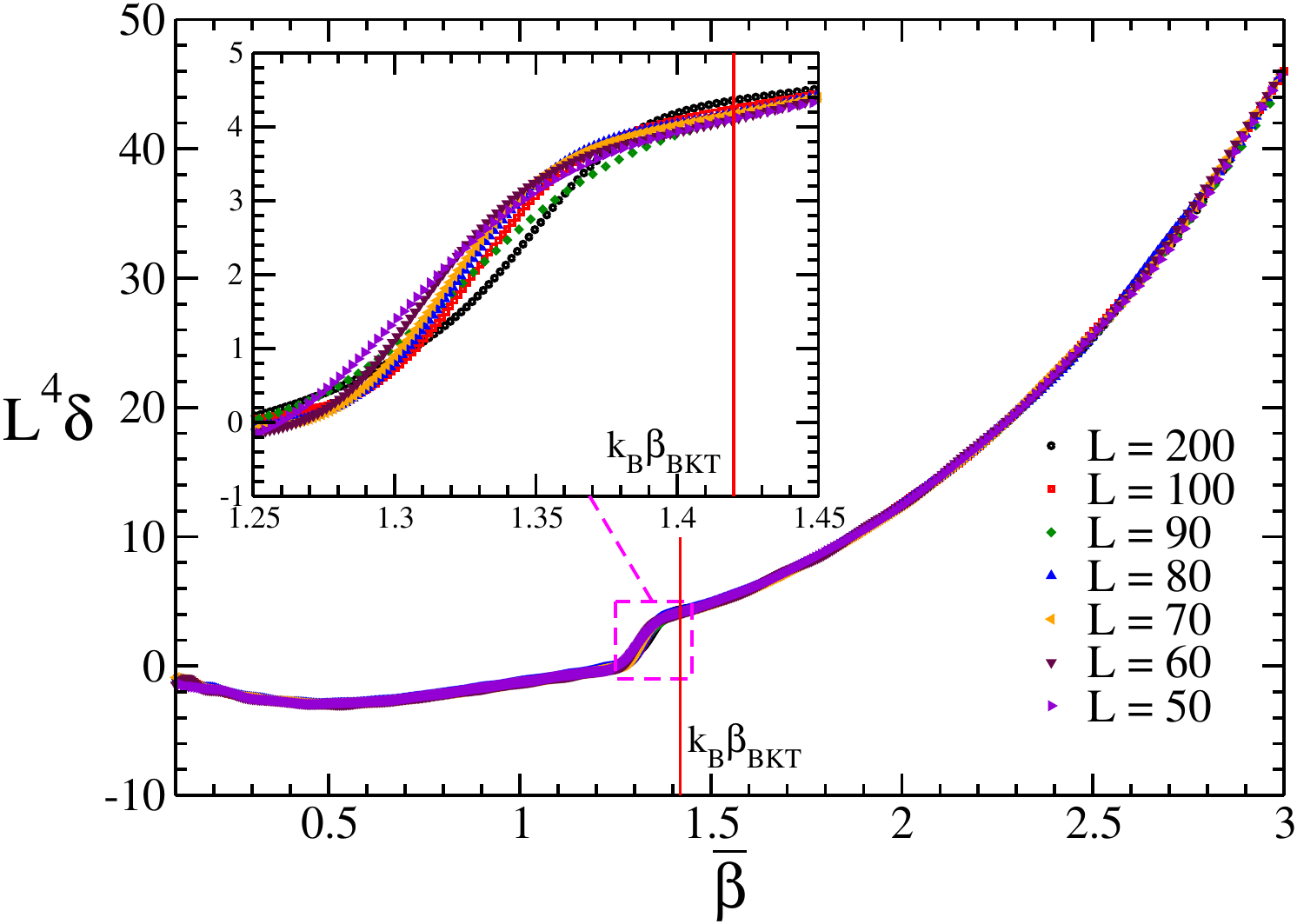} \\
(b) 
    \end{tabular}
    \caption{(Color online) The parametric microcanonical inflection point analysis for the XY model. In panel (a), we show $\gamma (\overline{\beta})$ and in panel (b) $\delta(\overline{\beta})$, for a linear system size $L=50,\ 60,\ 70,\ 80,\ 90, 100,\ \text{and } 200$. The vertical red solid line indicates the inverse temperature at the BKT transition for the XY model estimated from the zeros map. \label{microXY}}
\end{figure}

\begin{figure}[hbt!]
\centering
\includegraphics[width=0.475\textwidth,keepaspectratio=true,clip]{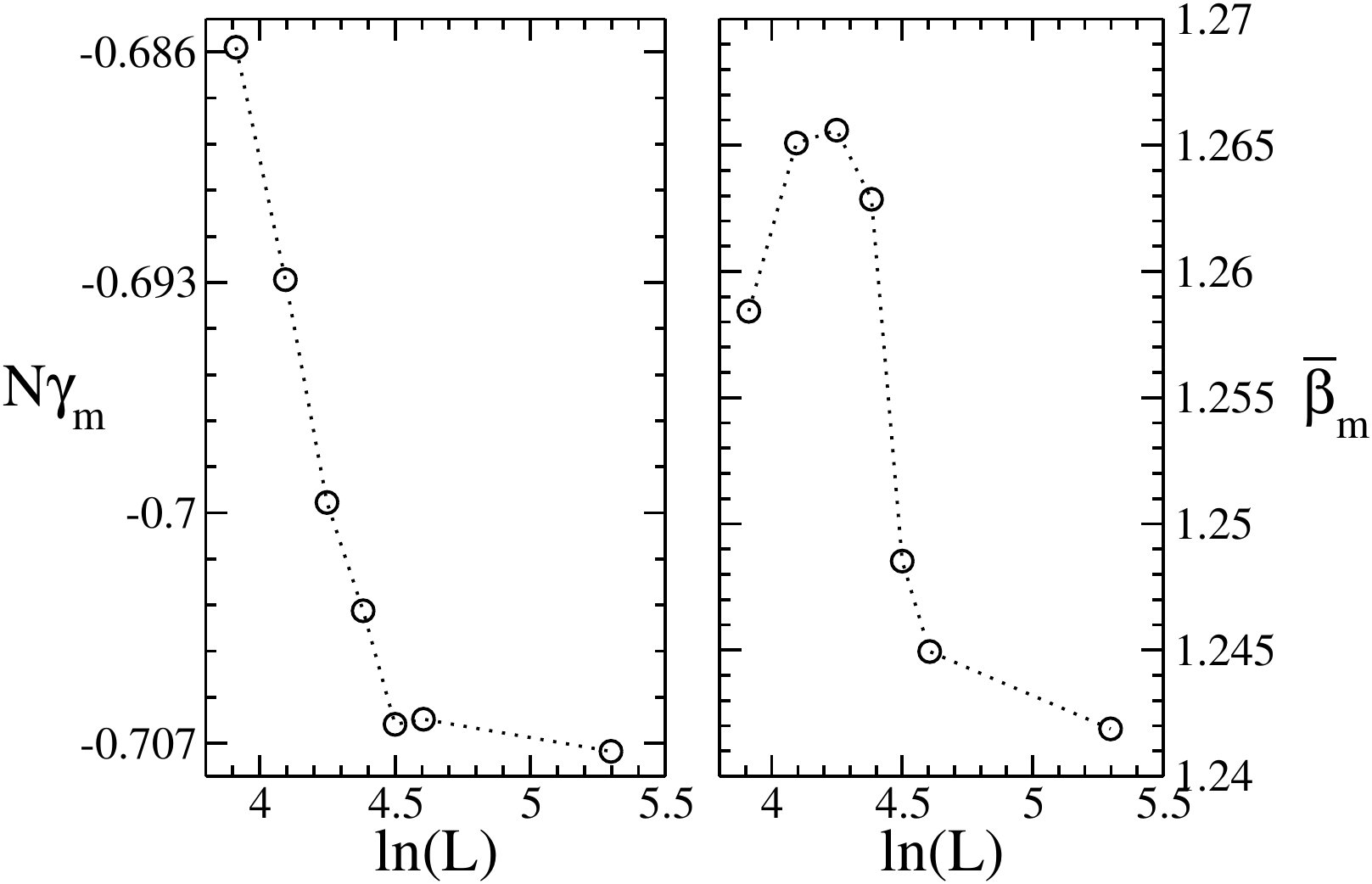} 
    \caption{Maximum of $N\gamma$ (left panel) and inverse temperature at peak position (right panel) as functions of the logarithm of linear system size for the XY-model.
        \label{fss_micro_xy}}
\end{figure}

In Fig.~\ref{microXY}, we present the parametric microcanonical inflection point analysis for this model, where five independent simulations were used to estimate the mean values and error bars. Given that finite-order derivatives of the free energy remain finite and continuous, one can anticipate no indication of a transition in these graphs. Panel (a) of the figure displays $L^2\gamma(\bar{\beta})$, where a local maximum with a negative value is observed. However, this feature does not exhibit any discernible scaling behavior, see Fig.~\ref{fss_micro_xy}, precluding the identification of a critical point, as expected. Panel (b) illustrates $L^4\delta(\bar{\beta})$, which similarly shows no evidence of a higher-order transition. Higher-order derivatives are not displayed due to the substantial error caused by the sensitivity of numerical derivatives to statistical fluctuations.
\noindent
\noindent

\noindent
\noindent
%======================================================================================================
\subsection{\label{resultsZeeman} Zeeman Model}

We chose to study the Zeeman model as an instance of a system that does not exhibit a phase transition at any finite temperature. This model consists of $N$ non interacting $1/2$-spins particles in a magnetic field, $\vec{B}$. The Hamiltonian can be expressed as
\begin{equation}
    \mathcal{H} = -\mu_B B \sum_{i=1}^N \sigma_i,
    \label{HamiltonianZeeman}
\end{equation}
where we employ reduced units such that $\mu_B=1$ represents the Bohr magneton, $|\sigma| = 1$ denotes the spin magnitude, and $\sigma_i=\pm 1$. Furthermore, $B = |\vec{B}|$ signifies the strength of the magnetic field, and the Boltzmann constant is set to $k_B=1$.

The exact number of states with energy $E$ for this model is well-known, as detailed in reference~\cite{salinas}. It is given by
\begin{align}
    \Omega(E,N) = \binom{N}{n} &= \frac{N!}{n!(N-n)!}, \nonumber \\
    & = \frac{N!}{\left[\frac{1}{2} \left(N - \frac{E}{B}\right)\right]!\left[\frac{1}{2} \left(N + \frac{E}{B}\right)\right]!},
    \label{DOSZeeman}
\end{align}
where $n$ represents the number of spins aligned with the magnetic field, and consequently, $(N-n)$ is the number of spins anti-aligned with $\vec{B}$. By directly evaluating the Hamiltonian, the energy density can be expressed as $\mathsf{e} = -Bm =-B (2n/N-1)$, where $m=M/N$ is the magnetization per spin.

The partition function can be written as~\cite{salinas}
\begin{align} \nonumber
    Z(\mathcal{B},N) & = e^{-\mathcal{B} B N}\sum_{n=0}^N \binom{N}{n}\Big( e^{2\mathcal{B} B}\Big)^n, \\ \label{zzeeman1}
    &= e^{-\mathcal{B} B N}\Big(1 + e^{2\mathcal{B} B} \Big)^N, \\ \label{zzeeman2}
    &=\Big[2\cosh(\mathcal{B} B)\Big]^N.
\end{align}
Thus, by inspecting Eq.~(\ref{zzeeman1}), $Z=0$ if $ e^{2 \mathcal{B} B} = -1$. Thus, $2 \mathcal{B} B = \pm i(2k-1)\pi$, for $k=1,2,\cdots $. This leads to $\beta_k = 0$ and
\begin{equation}
    \tau_k = \pm \frac{(2k-1)\pi}{2 B},
    \label{imagPartZerosZeeman}
\end{equation}
i.e., the Fisher zeros are evenly distributed along the imaginary inverse temperature axis. This analysis explains the presence of a vertical line pattern of zeros near the imaginary axis on the $\mathcal{B}$ maps (which correspond to a unit circle in the complex-$x$ map) for the LJ cluster and XY model, interpreted as a transition at infinite temperature. However, a perfect alignment with $\beta = 0$  is not observed in our results due to the exclusion of positive energy values ($E>0$)  from our simulations.

Furthermore, by considering Stirling's approximation, which states that $\ln{y!} = y\ln{y} - y + \mathcal{O}(\ln{y})$, the parametric curve for $N{\gamma}$ can be deduced to be
\begin{equation}
    N{\gamma}(\bar{\beta}) \approx -\left[\frac{1}{B}\cosh{\left(\bar{\beta} B\right)}\right]^2.
    \label{gammaBetaZeeman}
\end{equation}
This is a function with a negatively valued maximum at $\bar{\beta} =0$, also signaling the absence of a transition at finite temperature for this model, then corroborating the proposed methodology. Moreover, this peak approaches zero only as $B\to \infty$. 

By comparing with Eq.~(\ref{zzeeman2}), it can be stated that the analytic continuation of Eq.~(\ref{gammaBetaZeeman}) exhibits the same zeros as the partition function. At least in this specific case, the Fisher zeros and the zeros of $\gamma$ coincide across the entire complex plane. However, for other models, this equivalence certainly holds only for the real positive zero, which arises exclusively in the thermodynamic limit. 

It is pertinent to note that the direct substitution of Eq.~(\ref{gammaBetaZeeman}) into Eq.~(\ref{micro_c}) yields the specific heat as:
\begin{equation}
 c = (\bar{\beta} B)^2 \sech^2{(\bar{\beta}B)}.
\end{equation}
This expression exhibits a maximum, characteristic of the well-known Schottky anomaly~\cite{gopal, blundell, Kim2014}, at  
$\bar{\beta} B \approx 1.2$, which satisfies the condition
$\bar{\beta} B = \coth{(\bar{\beta}B)}$.
This maximum arises at a temperature where the thermal energy optimally excites transitions between discrete energy states, resulting in a smooth and broad peak in the specific heat, fundamentally distinct from the sharp features associated with phase transitions.
%%%%%%%%%%%%%%%%%%%%%%%%%%%%%%%%%%%%%%%%%%%%%%%%%%%%%%%%%%%%%%%%%%%%%%%%%%%%%%%%%%%%%%%%%%%%%%%%%%%%%%%%%%
\section{\label{conclusions} Conclusions}
%%%%%%%%%%%%%%%%%%%%%%%%%%%%%%%%%%%%%%%%%%%%%%%%%%%%%%%%%%%%%%%%%%%%%%%%%%%%%%%%%%%%%%%%%%%%%%%%%%%%%%%%%%
In this study, we introduce a technique for analyzing phase transitions, utilizing microcanonical quantities with inverse temperature, $\bar{\beta}$, as a parameter. By examining the behavior of thermodynamic quantities, such as entropy and its derivatives, as functions of $\bar{\beta}$, we observe that each type of transition has its own characteristic behavior. To introduce the method, we have studied several models in well-known universality class.

For first-order transitions, the parametric entropy curve exhibits a characteristic ``Z" shape, allowing for an equal-area Maxwell construction. Concurrently, the parametric curve of the second derivative of entropy ($\gamma$) forms a loop, with the knot point indicating the transition temperature. This loop structure effectively captures the behavior associated with the first-order transition.

In contrast, for second-order transitions, the parametric analysis of $\gamma$ reveals a negative-valued peak, consistent with traditional microcanonical inflection point analysis.

We have applied this framework to several model systems, including the Lennard-Jones cluster, the Ising model, the XY model, and the Zeeman model, demonstrating its effectiveness in characterizing first-order and second-order. Hence, the proposed method offers a powerful tool for understanding and classifying phase transitions in diverse physical systems.

Furthermore, we have explored the relationship between Fisher zeros and the parametric microcanonical curves, providing valuable insights into the underlying thermodynamic behavior. Specifically, we present a simplified demonstration of the zeros pattern in the unstable entropy region as a vertical line of equidistant zeros in the complex inverse temperature plane. Moreover, we demonstrate that the latent heat exhibits an inverse proportionality to the spacing between the zeros proximal to the real axis within this pattern.

As a perspective, we highlight the potential of the proposed methodology to serve as a refined and effective analytical tool for the classification of weak first-order transitions. The characterization of these transitions presents considerable analytical challenges, as they can be readily misidentified as second-order transitions due to their subtle signatures. Our primary results suggest that the proposed protocol can elucidate the pre transitional pseudocritical behavior at $\beta^* < \beta_{tr}$, observed in weak first-order transitions at $\beta_{tr}$~\cite{fernandez1992}, thereby offering a potential resolution to the aforementioned challenge.

\section*{Acknowledgments}
We would like to acknowledge helpful conversations with Dr. Michael Bachmann and Dr. Bruno B. Rodrigues.

This work received public financial support from Conselho Nacional de Desenvolvimento Científico e Tecnológico (CNPq), Brazil, under Grant No. 409719/2023-4.

 The authors have no competing interests to declare that are relevant to the content of this article.

%%%%%%%%%%%%%%%%%%%%%%%%%%%%%%%%%%%%%%%%%%%%%%%%%%%%%%%%%%%%%%%%%%%%%%%%%%%%%%%
\section*{Data Availability}

The data that support the findings of this article are openly available on Zenodo \cite{LJ_zenodo,XY_zenodo}.
%%%%%%%%%%%%%%%%%%%%%%%%%%%%%%%%%%%%%%%%%%%%%%%%%%%%%%%%%%%%%%%%%%%%%%%%%%%%%%%%%%%%%%%%%%%%%%%%%%%%%%%%%%
%\section*{References}

\bibliographystyle{unsrt}  
\bibliography{parametric_micro}

\end{document}